\documentclass[aps,prb,reprint,showpacs]{revtex4-2}
\usepackage{epsfig}\usepackage{subfigure}
\usepackage{dcolumn}
\usepackage{graphicx,graphics}
\usepackage{amsmath}
\usepackage{amssymb}
\usepackage{bm}
\usepackage{color}
\usepackage[utf8]{inputenc}
\usepackage{pifont}
\usepackage[colorlinks,linkcolor={blue},citecolor={blue},urlcolor={blue}]{hyperref}
\usepackage{mathtools}
\usepackage{mathtools}
\usepackage{bigints}
\newcommand{\Giret}{ {\rm G}^{{\rm R}}}
\newcommand{\Giadv}{ {\rm G}^{{\rm A}}}

\newcommand{\Giless}{ {\rm G}^{ <}}
\newcommand{\Sless}{ \Sigma^{ <}}

\newcommand{\Sigl}{ \Sigma^{ <}}

\newcommand{\bk}{{\bm k}}
\newcommand{\bq}{{\bm q}}

\newcommand{\ms}{m^{*}}
\newcommand{\omb}{\tilde{\omega}}

\newcommand{\homg}{\hbar \Omega}

\newcommand{\gra}{ g^{{\rm R/A}}}
\newcommand{\mvk}{\mathcal{V}_{1}k}
\newcommand{\fn}[1]{f(\omega+{#1}\homg)}
\newcommand{\fne}[1]{f(\epsilon_{#1,\bk})}
\newcommand{\En}[2]{(\epsilon_{{#1},\bk}-\omega)^{#2}}
\newcommand{\Ene}[1]{(\epsilon_{{#1},\bk}-\omega)}
\newcommand{\del}[1]{\delta(\epsilon_{{#1},\bk}-\omega)}
\newcommand{\fnt}[1]{f[\omega+{#1}\homg]}

\newcommand{\Ennq}[3]{(\epsilon_{{#1},\bk {#2}}-\epsilon_{{#3},\bk})}
\newcommand{\ma}{\mathcal{M}}
\newcommand{\mb}{\mathcal{B}}
\newcommand{\mc}{\mathcal{C}}

\begin{document}

\author{Mahmoud M. Asmar}
\affiliation{Department of Physics and Astronomy, The University of Alabama, Tuscaloosa, AL 35487, USA }
\author{Wang-Kong Tse }
\affiliation{Department of Physics and Astronomy, The University of Alabama, Tuscaloosa, AL 35487, USA }

\title{Impurity Screening and Friedel Oscillations in Floquet-driven Two-dimensional Metals}
\date\today

\begin{abstract}
  We develop a theory for the non-equilibrium screening of a charged impurity in a two-dimensional electron system under a strong time-periodic drive. Our analysis of the time-averaged polarization function and dielectric function reveals that Floquet driving modifies the screened impurity potential in two main regimes. In the weak drive regime, the time-averaged screened potential exhibits unconventional Friedel oscillations with multiple spatial periods contributed by a principal period modulated by higher-order periods, which are due to the emergence of additional Kohn anomalies in the polarization function. In the strong drive regime, the time-averaged impurity potential becomes almost unscreened and does not exhibit Friedel oscillations. This tunable Friedel oscillations is a result of the dynamic gating effect of the time-dependent driving field on the two-dimensional electron system.
\end{abstract}

\maketitle
\section{Introduction}

Periodically driving a solid by strong irradiation renormalizes its equilibrium band structure non-perturbatively, allowing on-demand control of the properties of the irradiated system through the laser intensity and frequency. This approach has been dubbed Floquet engineering~\cite{flreview1,flreview2} and has generated considerable recent interest in the properties of Floquet-driven systems, such as tunneling currents~\cite{Platero,FloqTunn4,FloqTunn1,FloqTunn2,FloqTunn3}, indirect magnetic exchange interaction~\cite{rkk1,rkk2}, transport~\cite{mitraandoka1,ftrans1,ftrans2,ftrans3,torres1,torres2,torres3,barderson,Floqtrans1,Floqtrans2} and optical response~\cite{opt1,Floqtrans3,opt2}. In addition to the dynamic control of material properties, Floquet-driven systems can exhibit nontrivial topological phases~\cite{Floqtop1,Floqtop3,Floqtop2,Floqtop4}.

Screening is a fundamental many-body effect of electron systems that
crucially determines many other physical properties. Screening
effects, for example, play an important role in transport which is
limited by the screened impurity potential in a disordered
sample. Transport in Floquet-driven systems have been largely
studied~\cite{mitraandoka1,ftrans1,ftrans2,ftrans3,torres1,torres2,torres3,barderson}
in the ballistic regime where scattering with screened external
impurities is absent. In general, consideration of time-dependent
screening in  non-equilibrium situations is a formidable many-body
problem requiring a numerical solution of both the Dyson's equation
and the quantum kinetic equation~\cite{Jauhobook}. For a periodically
driven system in the steady state, Floquet's theorem enables
additional simplifications of this problem. In the classic problem of
a charged external impurity in an electron gas, the effects of
screening are manifested through a charge density modulation around
the impurity known as the Friedel
oscillations~\cite{Friedel}. There have been
extensive theoretical studies on Friedel oscillations, and recent
interests have focused on low-dimensional systems such as
semiconductor structures~\cite{Fabian}, graphene~\cite{Falko,Wunsch,HwangFriedel} and other two-dimensional
materials~\cite{Agarwal,Chang}. Friedel oscillations provide
a promising probe into the unusual electronic properties of these
materials. A curious question that has been largely unexplored is the
influence of Floquet driving on the impurity screening and the Friedel
oscillations of a Floquet-driven metal. The
impurity-induced density oscillations could provide a useful signature
of the non-equilibrium dynamics of the driven electrons.

With this motivation, we theoretically formulate and study the screening response of an external impurity in a two-dimensional electron system (2DES)  driven by a monochromatic circularly polarized light. In the presence of a Floquet drive, the screening charges are driven out of equilibrium and non-equilibrium screening features emerge in the light-driven Friedel oscillations. In this work, we focus on the non-equilibrium steady state regime and present the standard random phase approximation (RPA) in the Floquet representation, and then calculate the time-averaged non-equilibrium polarization function, inverse dielectric function and screened potentials up to second order in the Floquet driving strength. Our results reveal that the screening of the irradiated 2DES operate in two distinct regimes, depending on the position of the Fermi level relative to the light-induced Floquet band shift. When the induced band shift is smaller than the Fermi level, screening effectiveness is reduced on average with increasing Floquet drive, and the screened impurity potential exhibits Friedel oscillations that consist of a principal oscillation and secondary oscillations with different periods that increase with the driving strength. The time-averaged polarization function exhibits new non-equilibrium Kohn anomalies whose values depend on the driving strength. When the induced band shift becomes larger than the Fermi level, there is a considerable reduction in screening and the electron gas becomes very weakly screened to almost unscreened. The impurity potential does not oscillate and remains almost bare-like except with minor spatial modulations. This dramatic difference in the screening response should be observable by scanning tunneling microscopy (STM)~\cite{stm1,stm2,stm3} by tuning the driving laser field.

Our paper is organized as follows. In Sec.~\ref{sec2} we introduce the problem of a driven 2DES with an embedded impurity, and we find the time-dependent RPA equations that determine the two-time dependent quantities governing the screening in this system, {\it i.e.}, the polarization function, dielectric function, and screened potential,  Sec.~\ref{sec2a}. We then assume that the system is periodically driven in Sec.~\ref{sec2b} which allows us to use the Floquet theory and derive a set of compact self-consistent RPA-Floquet matrix equations in the frequency space of the screening observables. In Sec.~\ref{sec2c}, we proceed to use the Keldysh-Floquet formalism and provide the formal expression for the free-particle polarization function of a periodically driven system in terms of the Keldysh-Floquet Green's functions. In Sec.~\ref{sec3}, we find the Floquet Hamiltonian of an irradiated 2DES, Sec.~\ref{sec3a}, which enables the analytical treatment of the Floquet Green's and free-particle polarization functions within an approximate scheme that relies on the strength of the systems driving, Sec.~\ref{sec3b}. We compare our fully numerical results of the time-averaged free-particle polarization to its approximate counterpart in Sec.~\ref{sec3c}, and we gain valuable insights into irradiation effects on the Kohn anomalies of the driven 2DES. We present and provide the physical interpretation of our analytical results for the time-averaged inverse dielectric function and the real space representation of the screened potential in Secs.~\ref{sec3d} and~\ref{sec3e}.

\section{Formalism}\label{sec2}
To establish the notations and framework, in this section we first review the standard random phase approximation (RPA) for screening in an interacting electron system that is not necessarily in equilibrium. Throughout this work, we assume translational symmetry in our system and two-particle quantities such as the polarization function depend only on spatial coordinate difference ${\bm r_{1}}-{\bm r_{2}}$ in the real space, or equivalently momentum $\bq$ in the reciprocal space. When the system is driven strongly out of equilibrium, time translational symmetry is broken and the system's response function no longer depends simply on the time difference $t_{1}-t_{2}$. Therefore, in the following we present the RPA formalism in the time domain and quantities such as the polarization function $\Pi_{{\rm sc}}$, dielectric function $\epsilon$, and screened potential $V_{{\rm sc}}$ depend on independent time arguments $t_{1}$ and $t_{2}$, {\it i.e.}, $\Pi_{{\rm sc}}(\bq;t_{1},t_{2})$, $\epsilon(\bq; t_{1},t_{2})$, and $V_{{\rm sc}}({\bm q};t_{1},t_{2})$. The time-dependent RPA equations we present are then specialized to the case for a time-periodic system, which enables the use of the Floquet formalism and the RPA equations to be expressed algebraically in convenient matrix forms similar to those in equilibrium.

\subsection{Time-dependent Polarization and Dielectric Functions }\label{sec2a}
We start by considering an external test charge on a metallic two-dimensional electron system (2DES). Electrons in the 2DES experience not only the bare {\it external potential} $\Phi_{{\rm ext}}(\bm r,t)$ due to the test charge, but also the {\it induced potential} $\Phi_{{\rm ind}}(\bm r, t)$ due to the screening charge density $n(\bm r, t)$ induced by the test charge in the electron gas. Within RPA, electrons are assumed to respond to the total {\it screened potential} $\Phi_{{\rm sc}}(\bm r, t)=\Phi_{{\rm ext}}(\bm r, t)+\Phi_{{\rm ind}}(\bm r,t)$~\cite{pines1, pines2,pines3,Cohen, pines}.
For a 2DES the charges are constrained on the plane and therefore $\rho(\bm r, z,t)=\rho(\bm r,t)\delta(z)=-e n(\bm r,t)\delta(z)$, where $\rho(\bm r,t)=-e n(\bm r,t)$, $e>0$ is the electron's charge, $r=\sqrt{x^2+y^2}$. The Poisson's equation gives the potential induced by the screening charges
\begin{equation}\label{p1}
 \nabla^{2} \Phi_{{\rm ind}}(\bm r,z,t)=\frac{4\pi e }{\epsilon_{b}}n(\bm r,t)\delta(z)\;.
\end{equation}
where $\epsilon_{b}$ is the background dielectric function. Eq.~\eqref{p1} can be solved easily in the $\bq$-space to give the induced potential on the plane $z = 0$, {\it i.e.} $\Phi_{{\rm ind}}(\bq,z=0,t)=\Phi_{{\rm ind}}(\bq,t)$ (from now on, we suppress the $z$-dependence for clarity)
\begin{equation}\label{poissoneq}
  \Phi_{{\rm ind}}(\bq,t)=\Phi_{\rm C}(\bq)n(\bq,t)\;.
\end{equation}
where $\Phi_{\rm C}(\bq)=-2\pi e/(\bq \epsilon_b)$ is the Fourier component of the bare Coulomb potential. Within RPA, the induced density response to the screened (total) potential is given by the non-interacting polarization function,
\begin{equation}\label{ind1}
  \rho(\bm r,t )=-e^{2}\int d\bm r' \int dt' \Pi(\bm r,t; \bm r', t') \Phi_{{\rm sc}}(\bm r', t')\;.
\end{equation}
Using $\rho(\bm r,t )=-e n(\bm r,t )$ to combine Eq.~\eqref{ind1} and Eq.~\eqref{poissoneq} in the momentum space yields
\begin{equation} \label{Phi_ind}
\Phi_{{\rm ind}}(\bq,t)= e\Phi_{\rm C}(\bq)\int dt' \Pi(\bq; t, t')  \Phi_{{\rm sc}}(\bq, t')\;.
\end{equation}
Substituting $\Phi_{{\rm ind}}$ from the above in $\Phi_{{\rm sc}}=\Phi_{{\rm ext}}+\Phi_{{\rm ind}}$ gives the
external potential in terms of the screened potential,
\begin{equation}\label{ext1}
\Phi_{{\rm ext}}(\bq,t)= \int dt'\left[\delta(t-t')- e\Phi_{\rm C}(\bq)\Pi(\bq; t, t')\right]  \Phi_{{\rm sc}}(\bq, t').
\end{equation}
The above equation immediately gives the dielectric function $\epsilon(\bq;t,t') = \delta \Phi_{{\rm ext}}(\bq,t)/\delta\Phi_{{\rm sc}}(\bq, t')$ in the linear screening regime
\begin{equation}\label{eps1}
\epsilon(\bq; t,t') =\delta(t-t')+ v_{\rm C}(\bq)\Pi(\bq; t, t')\;,
\end{equation}
where $v_{\rm C}(\bq)=-e\Phi_{\rm C}(\bq)=(2\pi e^{2})/(\bq \epsilon_{b})$ is the Coulomb potential energy. Moreover, the induced charge density in response to the external potential is given by the interacting polarization function $\Pi_{{\rm sc}}$
\begin{equation}\label{ind123}
  \rho(\bm r,t )=-e^{2}\int d\bm r' \int dt' \Pi_{{\rm sc}}(\bm r,t; \bm r', t') \Phi_{{\rm ext}}(\bm r', t')\;.
\end{equation}
Using the above with Eq.~\eqref{poissoneq} and $\Phi_{{\rm sc}} =\Phi_{{\rm ext}}+\Phi_{{\rm ind}}$ gives the Fourier component of $\Phi_{{\rm sc}}$,
\begin{equation}\label{sc1}
 \Phi_{{\rm sc}}(\bq, t)=\int dt' \left[\delta(t-t')+e\Phi_{\rm C}(\bq) \Pi_{{\rm sc}}(\bq,t,t')\right] \Phi_{{\rm ext}}(\bq, t')\;.
\end{equation}
Therefore the inverse dielectric function $\epsilon^{-1}(\bq;t,t') = \delta \Phi_{{\rm sc}}(\bq,t)/\delta\Phi_{{\rm ext}}(\bq, t')$ is related to the interacting polarization function as
\begin{equation}\label{eps2}
\epsilon^{-1}(\bq; t,t') =\delta(t-t')- v_{\rm C}(\bq)\Pi_{{\rm sc}}(\bq; t, t')\;.
\end{equation}

From the identity
\begin{equation}
  \int dt'' \epsilon(\bq,t,t'')\epsilon^{-1}(\bq,t'',t')=\delta(t-t')\;,
\end{equation}
one can derive the self-consistent equation for the interacting polarization function. Substituting Eq.~\eqref{eps1} for $\epsilon$ and Eq.~\eqref{eps2} for $\epsilon^{-1}$ into the above and performing the $t''$ integration, we recover the self-consistent RPA equation for the interacting polarization function
\begin{eqnarray}\label{piselfconst}
 \Pi_{{\rm sc}}&&(\bq,t,t')=\Pi(\bq,t,t')\\
 &&-\int dt'' dt''' \Pi(\bq,t,t'')v_{\rm C}(\bq, t'',t''') \Pi_{{\rm sc}}(\bq,t''',t'),\nonumber
\end{eqnarray}
where we have defined $v_{\rm C}(\bq, t,t')=v_{\rm C}(\bq)\delta(t-t')$. In addition, Eq.~\eqref{Phi_ind} gives the RPA equation for the screened electron-electron Coulomb interaction
\begin{eqnarray}\label{screendCol}
 V_{{\rm sc}}&&(\bq,t,t')=v_{\rm C}(\bq, t,t')\\
 &&-\int dt'' dt'''  v_{\rm C}(\bq, t,t'')\Pi(\bq,t'',t''') V_{sc}(\bq,t''',t')\;,\nonumber
\end{eqnarray}
where $ V_{{\rm sc}}(\bq,t,t')$$=$$-e\Phi_{{\rm sc}}(\bq,t,t')$, and $v_{\rm C}(\bq, t,t')$ is given in Eq.~\eqref{piselfconst}.

We have presented the two-time interacting polarization function, screened Coulomb interaction, and the dielectric and inverse dielectric functions for a system driven out of equilibrium. In what follows, we focus on systems subjected to a time-periodic drive and obtain the frequency-space representation of their screening properties.

\subsection{Frequency-Space Representation of RPA Equations}\label{sec2b}
In this section, we consider periodically driven systems for which the Floquet theory applies, and we find the frequency-space representation of the RPA equations.
Instead of independent time arguments $t$ and $t'$, a general two-time quantity $\mathcal{M}$ can be equivalently expressed in terms of the relative time $\tau=t-t'$ and the average time $T=(t+t')/2$. These correspond to microscopic and macroscopic time scales respectively, and it is conventional to perform a Fourier transformation of $\mathcal{M}(\tau,T)$ with respect to $\tau$ to obtain the Wigner representation $\mathcal{M}(\omega,T)$.
Recasting Eq.~\eqref{eps1} in terms of the relative and average times,
$
\epsilon(\bq; \tau,T) =\delta(\tau)+ v_{\rm C}(\bq)\Pi(\bq; \tau, T)
$,
and performing the Fourier transform for the relative time coordinates gives
$
\epsilon(\bq; \omega,T) =1+ v_{\rm C}(\bq)\Pi(\bq; \omega, T)
$.
Similarly, Eq.~\eqref{eps2} becomes
$
\epsilon^{-1}(\bq; \omega,T) =1- v_{\rm C}(\bq)\Pi_{\rm sc}(\bq; \omega,T)
$.
For a driving field with an arbitrary time dependence Eqs.~\eqref{piselfconst} and~\eqref{screendCol} do not generally yield a closed analytical form in terms of $\omega$ and $T$ as the above since their calculation requires the Moyal expansion~\cite{flreviewOKA,flreviewOKA2}.

On the other hand, if the driving field is time-periodic with a period $\mathcal{T}=2\pi/\Omega$ and frequency $\Omega$, our ability to use the Floquet theory allows for closed analytical forms of the $\omega$-space representation of not only the dielectric function and inverse dielectric function, but also the polarization function and screened Coulomb potential. The Floquet representation of a periodic two-time function and their convolutions are given in Eqs.~\eqref{att}-\eqref{att4} in Appendix~\ref{apend1}. These transformations facilitate the expression of the screening equations [Eqs.~\eqref{eps1}, \eqref{eps2}, \eqref{piselfconst} and \eqref{screendCol}] in the extended-frequency Floquet representation.
The Floquet representation  of the dielectric function in Eq.~\eqref{eps1} is
\begin{equation}\label{eps1FRomegaext}
\left[\epsilon(\bq,\omega)\right]_{m,n} =\delta_{m,n}+ v_{\rm C}(\bq)\left[\Pi(\bq,\omega)\right]_{m,n}\;,
\end{equation}
similarly, for the inverse dielectric function Eq.~\eqref{eps2},
\begin{equation}\label{eps2FRomegaext}
\left[\epsilon^{-1}(\bq, \omega)\right]_{m,n} =\delta_{m,n}- v_{\rm C}(\bq)\left[\Pi_{{\rm sc}}(\bq,\omega)\right]_{{m,n}}\;,
\end{equation}
and the Floquet representation of the self-consistent equations for the interacting polarization function, Eq.~\eqref{piselfconst}, and the screened Coulomb potential, Eq.~\eqref{screendCol}, are
\begin{eqnarray}\label{piselfFloquetext}
&&\left[\Pi_{{\rm sc}}(\bq,\omega)\right]_{{m,n}}=\\
&&\left[\Pi(\bq,\omega)\right]_{m,n}  -v_{\rm C}(\bq) \sum_{j}\left[\Pi (\bq,\omega)\right]_{m,j}\left[\Pi_{{\rm sc}} (\bq,\omega)\right]_{j,n}.\nonumber
\end{eqnarray}
and
\begin{eqnarray}\label{screendColFloext}
&&\left[V_{{\rm sc}}(\bq,\omega)\right]_{{m,n}}=\\
&&v_{\rm C}(\bq)\delta_{m,n}-v_{\rm C}(\bq)\sum_{j}\left[\Pi(\bq,\omega)\right]_{{m,j}}\left[V_{{\rm sc}}(\bq,\omega)\right]_{j,n},\nonumber
\end{eqnarray}
where $m, n, j \in \{0,\pm 1, \pm2, \ldots \}$ are the Floquet indices and $\omega \in (-\infty,\infty)$ is the extended zone frequency. Converting the Floquet component forms of  Eqs.~\eqref{eps1FRomegaext}-\eqref{eps2FRomegaext} to their matrix representation results in a set of self-consistent matrix equations resembling the well-known RPA relations~\cite{pines},
\begin{subequations}\label{matforms}
  \begin{eqnarray}
   &&\epsilon(\bq,\omega) =\mathbb{I}+ v_{\rm C}(\bq)\Pi(\bq,\omega), \\
    &&\epsilon^{-1}(\bq, \omega) =\mathbb{I}- v_{\rm C}(\bq)\Pi_{{\rm sc}}(\bq,\omega), \\
    &&\Pi_{{\rm sc}} (\bq,\omega) =\left[\mathbb{I}+v_{\rm C}(\bq)\Pi (\bq,\omega)\right]^{-1}\Pi (\bq,\omega), \\
  &&V_{{\rm sc}}(\bq,\omega)=\left[\mathbb{I}+v_{\rm C}(\bq)\Pi (\bq,\omega)\right]^{-1}v_{\rm C}(\bq),
  \end{eqnarray}
\end{subequations}
where $\epsilon(\bq,\omega)$, $\Pi_{\rm sc} (\bq,\omega)$ $[\Pi(\bq,\omega)]$, and $V_{{\rm sc}}(\bq,\omega)$, are the Floquet matrix representations of the dielectric function, interacting (free-particle) polarization function, and screened Coulomb potential, respectively, and $\mathbb{I}$ is the identity matrix. The set of  Eqs.~\eqref{matforms} comprise the Floquet-RPA equations and encode the non-equilibrium screening effects under a time-periodic driving into a convenient matrix form.

For a periodically driven system in the steady state, we are interested in its average response over time and therefore it is convenient to define the one-cycle average of the system's physical observables. It can be readily established that the time average of a periodic two-time function is given by its  $(0,0)$-Floquet component (for derivation, see Appendix~\ref{appen2}). Hence, the time-averaged $\epsilon(\bq,\omega)$, $\epsilon^{-1}(\bq, \omega)$, $\Pi_{\rm sc} (\bq,\omega)$, and $V_{{\rm sc}}(\bq,\omega)$, are given by the $(0,0)$-Floquet components of their matrix equations in Eq.~\eqref{matforms}.

This section has provided the Floquet representation of the RPA screening equations in a periodically driven system, Eq.~\eqref{matforms}. The central quantity that enters into Eq.~\eqref{matforms}  is the free-particle polarization function $\Pi$, which will be derived in the next section.

\subsection{Free-particle Polarization Function}\label{sec2c}
We will now proceed with the derivation of the non-interacting polarization function $\Pi$. By introducing a weak external perturbing potential $V_{{\rm ext}}(\bm r, t)=-e\Phi_{{\rm ext}}({\bm r}, t)$ to  the 2DES,  $\Pi$ will be obtained as the response function of the induced electron density to the total potential in the system.
The total potential is given by $ V_{{\rm sc}}(\bm r, t)= V_{{\rm ext}}(\bm r, t)+V_{{\rm ind}}(\bm r, t)$, where $V_{{\rm ind}}(\bm r, t)$ is the induced potential due to the screening charges. The number density $n(\bm r,t)$ of the screening charges can be calculated from the lesser Green's function as
\begin{equation}\label{KF1}
 n(\bm r,t) =-i \lim_{\substack{{\bm r} \to {\bm r}'\\ t \to t'}}{\rm Tr}\left[\delta \Giless (\bm r,t;{\bm r}',t')\right]\;,
\end{equation}
where
the trace is over all the degrees of freedom including the spin.
Using the Dyson's equation within the non-equilibrium Green's function formalism the lesser Green's function $\delta\Giless$ can be obtained in linear response to the screened potential,
\begin{eqnarray}\label{LR1}
  &&\delta \Giless(\bm r,t;{\bm r}',t')=\\
  &&\int dt'' dr'' \bigg\{\Giret (\bm r, t; {\bm r}'', t'')V_{{\rm sc}}({\bm r}'', t'') \Giless ({\bm r}'', t''; {\bm r'}, t')\nonumber \\
  &&\;\;\;\;\;\;\;\;+\Giless(\bm r, t; \bm{r}'', t'') V_{{\rm sc}}({\bm r}'', t'') \Giadv ({\bm r}'', t''; r', t')\bigg\}\;,\nonumber
\end{eqnarray}
where $\Giret$ ($\Giadv$) is the retarded (advanced) Green's function. Thus the Fourier component of the electron density $n(\bq,t)$ follows from Eqs.~\eqref{KF1}-\eqref{LR1} as
\begin{eqnarray}\label{KF2}
  && n(\bq,t) =\\
  &&-2i\int\frac{d\bk}{(2\pi)^2}\int dt''\bigg\{ \Giret_{\bk+\bq}( t, t'')V_{{\rm sc}}(\bq,t'') \Giless_{\bk} (t'',t) \nonumber\\
 && \;\;\;\;\;\;\;\;\;\;\;+\Giless_{\bk}( t, t'') V_{{\rm sc}}(\bq,t'') \Giadv_{\bk -\bq} ( t'', t')\bigg\}\;.\nonumber
\end{eqnarray}
The non-interacting polarization function within RPA is given by the response of the induced density to the total screened potential $\Pi(\bq, t,t') = \delta n(\bq,t)/\delta V_{{\rm sc}}(\bq,t')$. From Eq.~\eqref{KF2} we therefore have
\begin{eqnarray}\label{pitime}
\Pi&&(\bq, t,t')=2i\int\frac{d\bk}{(2\pi)^2}\bigg\{\\
&&\Giret_{\bk+\bq}( t, t') \Giless_{\bk} (t',t)+\Giless_{\bk}( t, t') \Giadv_{\bk -\bq} ( t', t)\bigg\}\;.\nonumber
\end{eqnarray}
Using the Floquet representation for Green's functions~\cite{flreviewOKA,flreviewOKA2}, $\Giret_{\bk+\bq}( t, t'') \Giless_{\bk} (t'',t)$ can be expressed as
\begin{eqnarray}\label{endeq22}
&&\Giret_{\bk+\bq}( t, t'') \Giless_{\bk} (t'',t)=\sum_{s,l}\int_{-\infty}^{\infty}\frac{d \omega_{1}}{2\pi}\frac{d \omega_{2}}{2\pi}\\
&&\times e^{-i(\omega_{1}-\omega_{2}+(l-s)\Omega/2)\tau}  e^{-i(l+s)\Omega T} [\Giret_{\bk+\bq}( \omega_{1})]_{l,0}[ \Giless_{\bk}(\omega_{2})]_{s,0},\nonumber
\end{eqnarray}
where $\omega_{1,2}$ are extended-zone frequencies, $\tau=t-t''$ is the relative time, and $T=(t+t'')/2$ is the average time. Similarly, $\Giless_{\bk}( t, t'') \Giadv_{\bk -\bq} ( t'', t')$ satisfies Eq.~\eqref{endeq22} by replacing ${\rm R}$ by $ < $ and $<$ by ${\rm A}$. Substituting the transformations of $\Giret_{\bk+\bq}( t, t'') \Giless_{\bk} (t'',t)$ and $\Giless_{\bk}( t, t'') \Giadv_{\bk -\bq} ( t'', t')$ in Eq.~\eqref{pitime} and using the definitions in Appendix~\ref{apend1} [Eq.~\eqref{att3}], we find the $(m,n)$ Floquet matrix component of the free-particle polarization function,
\begin{eqnarray}\label{pi0final}
 && \left[\Pi(\bq,\omb)\right]_{m,n} =2i\int_{-\infty}^{\infty}\frac{d\omega}{2\pi}\bigg\{\\
 &&\sum_{p}\int_{-\infty}^{\infty}\frac{d\bk}{(2\pi)^2}{\big\{}[ \Giret_{\bk+\bq}(\omega+ \omb)]_{m,p}[\Giless_{\bk} ( \omega-n\Omega)]_{p,n}\nonumber\\
  &&\;\;\;\;\;\;\;\;\;\;\;\;\;\;\;\;\;\;+[\Giless_{\bk}( \omega+ \omb)]_{m,p} [\Giadv_{\bk -\bq} (\omega-n\Omega)]_{p,n}{\big\}}\bigg\}.\nonumber
\end{eqnarray}

We have established the Floquet-RPA formalism for periodically-driven systems, and we also have found the formal expression for their free particle polarization function matrix. In the following section, we implement our formalism in studying the effects of monochromatic irradiation on the screening properties of 2DESs.

\section{Impurity Screening and Friedel Oscillations in Periodically Driven 2DESs}~\label{sec3}

We now consider an external immobile  impurity embedded in the 2DES and study the effects of Floquet driving on the screening of the impurity potential. The impurity is assumed to be immobile and remains stationary in the presence of the driving field.
The potential due to the impurity is  $\Phi_{{\rm ext}}(\bq,t)=\Phi_{{\rm ext}}(\bq)=-2\pi Ze/(\epsilon_{b}\bq)$, where $e>0$ is the electron charge, $\epsilon_{b}$ is the background dielectric constant of the 2DES and $Z$ is the charge number.
Taking the time average of Eqs.~\eqref{sc1}-\eqref{eps2} gives the time-averaged screened potential of the impurity (see Appendix~\ref{appen3} for a derivation),
\begin{eqnarray}\label{screened2}
  \overline{V}_{{\rm sc}}(\bq)= \overline{\epsilon^{-1}}(\bq){\mathbb{V}}_{{\rm C}}(\bq)\;,
\end{eqnarray}
where ${\mathbb{V}}_{{\rm C}}(\bq)=e\Phi_{{\rm ext}}(\bq)$ and $\overline{\epsilon^{-1}}(\bq)=\left[\mathbb{I}+ v_{\rm C}(\bq)\Pi_{0}(\bq)\right]^{-1}_{0,0}$ is the time-averaged inverse dielectric function. Here we note that only the inverse dielectric function in the `static' limit [\textit{i.e.}, $\overline{\epsilon^{-1}}(\bq,0) \equiv \overline{\epsilon^{-1}}(\bq)$]
enters into the expression of the screened potential above due to the static nature of the impurity.
The time-averaged screened Coulomb interaction requires the Floquet matrix representation of the free-particle polarization function from Eq.~\eqref{pi0final} in the static limit $\Pi(\bq,0) \equiv  \Pi(\bq)$
\begin{eqnarray}\label{pi0finalnew}
 && [\Pi(\bq)]_{m,n} =\\
 &&2i\int_{-\infty}^{\infty}\frac{ d\omega}{2\pi}\sum_{p}\int_{-\infty}^{\infty}\frac{d\bk}{(2\pi)^2}{\big\{}[ \Giret_{\bk+\bq}(\omega)]_{m,p}[\Giless_{\bk} ( \omega-n\Omega)]_{p,n}\nonumber\\
  &&\;\;\;\;\;\;\;\;\;\;\;\;\;\;\;\;\;\;+[\Giless_{\bk}( \omega)]_{m,p} [\Giadv_{\bk -\bq} (\omega-n\Omega)]_{p,n}{\big\}}.\nonumber
\end{eqnarray}
The Floquet representation of the polarization function in Eq.~\eqref{pi0finalnew} requires that the driven system has the same periodicity as the driving field. The latter is realized when the system has fully relaxed into the non-equilibrium steady state (NESS). As the NESS is reached, the system is described by the Floquet theory, and its Green's functions naturally take their Floquet representation. In order to understand the formation of the NESS and the restoration of the driven system's periodicity, we assume that a two-dimensional metallic system is taken out of equilibrium via irradiation by a monochromatic light at time $t=0$. When the light is turned on, and the system is continuously driven,  energy is also continuously pumped into the metallic system. However, the various inelastic scattering processes present in a solid and the ambient neighbouring the solid provide several intrinsic and extrinsic channels of energy relaxation that keep the pumped system from heating up indefinitely. In our model, we adopt a specific method of energy dissipation. We assume that thermalization in the irradiated system is attained in the NESS through the connection to a fermion bath, {\it e.g.}, metallic leads. Hence, the lesser Green's function can be determined by the Keldysh-Floquet formalism, where
\begin{equation}\label{glessfin}
 \Giless_{\bk}(\omega)=\Giret_{\bk}(\omega)\Sless_{\bk}(\omega)\Giadv_{\bk}(\omega)\;,
\end{equation}
and $\Sigl_{\bk}(\omega)$ is obtained from integrating over the fermionic bath degrees of freedom. Considering the wide-band approximation for the bath ~\cite{FloqTunn4,rkk1,rkk2}, we have $[\Sigl(\omega)]_{n,m}=2i\eta f(\omega+n\hbar\Omega)\delta_{n,m}$, where $ f(\omega)$ is the Fermi distribution function and $\eta$ is a small broadening parameter. Additionally, the retarded and advanced Green's functions are given in terms the periodically driven system's Floquet Hamiltonian $H_{\rm F}$,
\begin{equation}\label{Gra}
  \left[{\rm G}_{\bk}^{{\rm R /A}}(\omega)\right]_{m,n}= \left\{\left(\omega\pm i\eta\right)\delta_{m,n} -\left[H_{{\rm F}}(\bk)\right]_{m,n}\right\}^{-1}\;.
\end{equation}
Hence, our knowledge of the Floquet Hamiltonian of a light-driven 2DES is essential for evaluating the free particle polarization function and the screened Coulomb potential. In what follows, we discuss the Floquet Hamiltonian of the driven system.

\subsection{Floquet Hamiltonian of Irradiated 2DESs}~\label{sec3a}

For the two-dimensional electron gas we consider the model of a single parabolic band Hamiltonian with spin degeneracy
$
H_{0}(\bk)=\hbar^2k^{2}/(2\ms)
$,
where $\ms$ is the effective electron mass. The 2DES is illuminated by a monochromatic laser beam perpendicularly incident onto the plane. The driving laser field enters into the Hamiltonian through the minimal coupling scheme and the time-dependent Schr\"{o}dinger equation describing the irradiated 2DES electrons is given by
\begin{equation}\label{tdepH}
\frac{\hbar^2}{2\ms}\left|\bk+(e/\hbar){\bm A}(t)\right|^2|\psi_{\alpha}(t)\rangle=i\hbar\partial_{t}|\psi_{\alpha}(t)\rangle\;,
\end{equation}
where $\bm{A}(t)=(E_{0}/\Omega)[\sin(\Omega t)\hat{x}+\sin(\Omega t+\varphi)]$
is the vector potential, $\varphi$ determines the light polarization,
$\Omega$ is the driving frequency and
$E_{0}$ is the driving electric field amplitude. For the rest of our
paper, it will be convenient to define $\mathcal{A}=eE_{0}/\hbar \Omega$ to indicate the
driving strength of the system.

Since the driven system is time periodic, the solutions for  Eq.~\eqref{tdepH} follows from Floquet's theorem as
$
|\psi_{\alpha}(t)\rangle=e^{-i\epsilon_{\alpha}\tau/\hbar}|\Phi_{\alpha}(\tau)\rangle$,
where $\epsilon_{\alpha}$ is called the quasienergy and $\alpha=(l,\bk)$ where $l$ is
the Floquet index and the $\bk$ momentum in our system. The quasienergies are
defined modulo $\hbar\Omega$ and the Floquet index $l$ classifies the
Floquet sidebands. $|\Phi_{l,\bk}(t)\rangle$
are known as the Floquet states, and possess the same periodicity as the
Hamiltonian so that $|\Phi_{l,\bk}(t+\mathcal{T})\rangle=|\Phi_{l,\bk}(t)\rangle$, where $\mathcal{T}=2\pi/\Omega$ denotes the driving period. Moreover, these states satisfy the Floquet eigenvalue equation
\begin{equation}\label{floquetEqn}
  H_{\rm F}(t,\bk)|\Phi_{l,\bk}(t)\rangle=\epsilon_{l,\bk}|\Phi_{l,\bk}(t)\rangle\;,
\end{equation}
where
$
H_{{\rm F}}(t,\bk)=H(t,\bk)-i\hbar\partial_{t},
$
is called the Floquet Hamiltonian. Since the Floquet states are
time periodic, they can be represented in a Fourier series
$
|\Phi_{l,\bk}(t)\rangle=\sum_{n=-\infty}^{\infty}e^{-in\Omega t}|\phi^{n}_{l,\bk}\rangle$,
where $|\phi^{n}_{l,\bk}\rangle$ are the Fourier components. Substituting the Fourier decomposition of the Floquet states in Eq.~\eqref{floquetEqn} yields a time-independent Floquet eigenvalue equation
\begin{equation}\label{timeindepFloq}
\sum_{n}{\left\{\left[H_{{\rm F}}(\bk)\right]_{m,n}-\delta_{m,n}m\hbar \Omega\right\}|\phi^{n}_{l,\bk}\rangle}=\epsilon_{l,\bk}|\phi^{m}_{l,\bk}\rangle
\end{equation}
where the Floquet matrix $[H_{\rm F}(\bk)]_{m,n}$ is
\begin{equation}\label{floquetmatrix}
\left[H(\bk)\right]_{m,n}=\frac{1}{T}\int_{0}^{T}{e^{i(m-n)\Omega t} H(t) dt}\;,
\end{equation}
and the Floquet Hamiltonian becomes
\begin{equation}
\left[H_{\rm F}(\bk)\right]_{m,n}=\left[H(\bk)\right]_{m,n}-\delta_{m,n}m\hbar\Omega\;.
\end{equation}
For our case, the Floquet Hamiltonian of the irradiated 2DES can be obtained as
\label{Hmn}
\begin{eqnarray}\label{Hmn}
[H_{{\rm F}}(\bk)&&]_{m,n}=\left(\frac{\hbar^{2}{k}^{2}}{2\ms}+\mathcal{V}_{2}-m\hbar\Omega\right)\delta_{m,n}\nonumber \\ &&-i\mathcal{V}_{1}\left[\delta_{m+1,n}(k_{x}+e^{i\varphi}k_{y})-\delta_{m-1,n}(k_{x}+e^{-i\varphi}k_{y})\right] \nonumber \\ &&-\mathcal{V}_{2}\frac{\cos(\varphi)}{2}\left[e^{-i\varphi}\delta_{m-2,n}+e^{i\varphi}\delta_{m+2,n}\right]\;,
\end{eqnarray}
where for convenience we have defined $\mathcal{V}_{1}$ and $\mathcal{V}_{2}$ through $2\mathcal{V}_{1}\ms/\hbar^2$$=$$\mathcal{A}$, and $2\mathcal{V}_{2}\ms/\hbar^2$$=$$\mathcal{A}^2$.  Since $\hbar \mathcal{A}=eE_{0}/\Omega$ has the dimension of momentum, $\mathcal{V}_{1}/\hbar$ and $\mathcal{V}_{2}$ are the characteristic velocity and the energy scale corresponding to the driving strength and light-induced band shift, respectively.

We shall focus on the case of a circularly polarized (CP) laser field $\varphi=\xi\pi/2$ ($\xi=\pm$ for left and right circular polarizations). The Floquet Hamiltonian Eq.~\eqref{Hmn} then takes the form
\begin{eqnarray}\label{HFCP}
  \left[H_{\rm F}(\bk)\right]_{m,n}&&=\left(\frac{\hbar^{2}k^{2}}{2\ms}+\mathcal{V}_{2}-m\hbar\Omega\right)\delta_{m,n}-\\
          &&i\mathcal{V}_{1}\left[\delta_{m+1,n}(k_{x}+i\xi k_{y})-\delta_{m-1,n}(k_{x}-i\xi k_{y})\right]. \nonumber
\end{eqnarray}
The quasienergy eigenvalues of the Floquet Hamiltonian in this case are exactly solvable and given by~\cite{Shirley,rkk1}
\begin{equation}\label{exactenergy}
\epsilon_{l,k}=\hbar^{2}k^{2}/{(2\ms)}+\mathcal{V}_{2}-l\hbar\Omega\;.
\end{equation}
The quasienergies consist of a ladder of parabolic dispersions displaced by integer multiples of the photon energy $\hbar\Omega$. Importantly, the zeroth order quasienergy band is displaced by an optically induced energy shift $\mathcal{V}_{2}$ dependent on the driving field amplitude and frequency. Thus the relative position of the quasienergy bands with respect to the 2DES's Fermi level can be controlled by tuning the driving field parameters.

By finding the Floquet Hamiltonian for a 2DES monochromatically irradiated with a CP light, we proceed to determine the Green's functions in Eqs.~\eqref{glessfin} and \eqref{Gra} which are essential for obtaining the free particle polarization function, Eq.~\eqref{pi0finalnew}, the time-averaged dielectric function, and the time-averaged screened Coulomb potential, Eq.~\eqref{screened2}.

\subsection{Analytical Treatment of the Free-Particle Floquet Polarization Function Matrix}~\label{sec3b}
The evaluation of Floquet Green's functions, in Eqs.~\eqref{glessfin} and \eqref{Gra}, and the non-interacting polarization function matrix elements is often a numerical task due to the infinite dimensionality of the Floquet Hamiltonian Eq.~\eqref{HFCP}. Nevertheless, the dimensionless parameter $\mathcal{V}_{1}k/\homg$, which characterizes the driving strength, constrains the number of relevant photon-number processes and, consequently, the relevant Floquet sidebands. The latter observation is crucial in the formulation of an approximation method that captures the analytical forms of the Green's function and free-particle polarization function of the driven system in successive powers of the light driving strength. In this section, we present the Floquet polarization function matrix up to second order in the driving strength to illustrate our approximation scheme. In Appendix~\ref{appen4}, we provide the approximate second order forms of $\Giret_{\bk}(\omega)$, $\Giadv_{\bk}(\omega)$, and $\Giless_{\bk}(\omega)$. Our formulation of the approximate Green's functions is carried out by expanding the Floquet Green's functions to second order in $\mathcal{V}_{1}k/\homg$, which has a pentadiagonal matrix structure.

Substituting the approximate expressions of $\Giless$, $\Giret$ and $\Giadv$ [Eqs.~\eqref{GRAexact} and~\eqref{glessapproximation}] into Eq.~\eqref{pi0finalnew} and integrating over $\omega$, we find that $\Pi(\bq)$ also takes a pentadiagonal matrix structure with components given by
\begin{eqnarray}\label{diagonalpi}
&&\left[\Pi(\bq)\right]_{n,n}=\frac{1}{2\pi^2}\int d^2\bk \left[ 1-2\left(\frac{\mathcal{V}_1}{\homg}\right)^2[k^2+q^2] \right]\\
&&\times\sum_{\alpha=\pm}{\frac{\fne{0}}{\Ennq{\alpha n}{+\alpha\bq}{0}}}+\frac{1}{2\pi^2}\left(\frac{\mathcal{V}_1}{\homg}\right)^2\bigg\{\int d^2\bk\bigg[\nonumber\\
&&\left.\sum_{\tau,\alpha=\pm}\frac{\fne{\tau} k^2}{\Ennq{\alpha n}{+\alpha\bq}{0}}+\frac{\fne{0}[q^2-k^2]}{\Ennq{\alpha n+\tau}{+\alpha\bq}{0}}\right]\bigg\}\nonumber
\end{eqnarray}
\begin{eqnarray}\label{pi2}
&&\left[\Pi(\bq)\right]_{n+1,n}=\left[\Pi(\bq)\right]^{*}_{n,n+1}=\frac{-i}{2\pi^2}\left(\frac{\mathcal{V}_{1}q}{\homg}\right)\bigg\{\\
&&\int d^2\bk \sum_{\alpha=\pm}\left[\frac{e^{- i\theta_{q}}\fne{0}}{\Ennq{\alpha n}{+\alpha \bq}{0}}-\frac{e^{- i\theta_{q}}\fne{0}}{\Ennq{\alpha n+\alpha}{+\bq}{0}}\right]\nonumber,
\end{eqnarray}
\begin{eqnarray}\label{pi3}
&&\left[\Pi(\bq)\right]_{n+2,n}=\left[\Pi(\bq)\right]^{*}_{n,n+2}=-\frac{1}{(2\pi)^2}\left(\frac{\mathcal{V}_{1}q}{\homg}\right)^2\bigg\{\nonumber\\
&&\int d^2\bk\sum_{\alpha=\pm} \left[\frac{e^{-2i\theta_{q}}\fne{0}}{\Ennq{\alpha n}{+\alpha \bq}{0}}+\frac{e^{- 2i\theta_{q}}\fne{0}}{\Ennq{\alpha n+2\alpha}{+\alpha \bq}{0}}\right.\nonumber\\
&&\;\;\;\;\;\;\;\;\;\left.-2\frac{e^{- 2i\theta_{q}}\fne{0}}{\Ennq{\alpha n+\alpha}{+\alpha \bq}{0}}\right],
\end{eqnarray}
It is interesting to notice that the matrix elements of $\Pi(\bq)$ are modified versions of the well known Lindhard form.

The free-particle polarization function is essential for finding the time-averaged quantities of interest, such as the time-averaged polarization function, inverse dielectric function, and the screened Coulomb potential. In the next section, we evaluate the time-averaged free-particle polarization function numerically, and we compare it to analytical results we obtained from our approximate method.

\subsection{Time-Averaged Free-Particle Polarization Function}~\label{sec3c}
\begin{figure}
  \centering
  \includegraphics[width=\columnwidth]{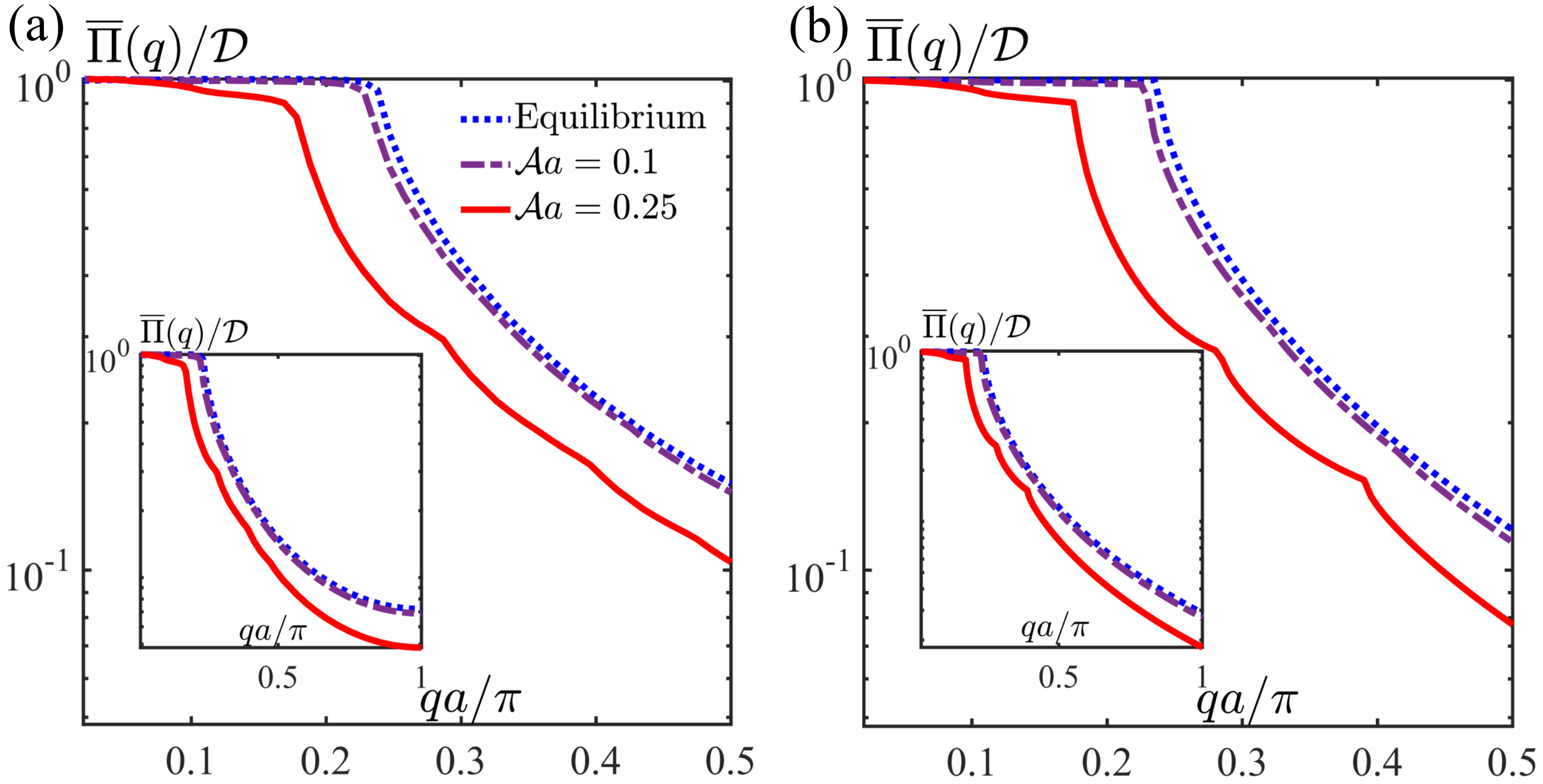}
  \caption{Time-averaged polarization function calculated (a)
    numerically, (b) analytically, in units of
    $\mathcal{D}=m^{*}/(\pi\hbar^2)$, for the dimensionless driving strengths
    $\mathcal{A}a=0.1,0.25$, and the equilibrium case. The driving
    frequency is taken as $\hbar\Omega=0.3$ eV and a field amplitude
    as $E_{0}=84,\;210$ MV/m. The 2DES Fermi level is taken at
    $E_{F}=140$ meV, its effective mass $m^{*}=0.25m_{0}$, $m_{0}$ is
    the bare electron mass, and $a=4$ $\rm \AA$.
   The insets in (a) and (b) show extended $q$-ranges of $\overline{\Pi}(q)$.}\label{fig1}
\end{figure}
In this section we evaluate analytically and numerically the time-averaged free-particle polarization function, which is given by the $(0,0)$-component of its Floquet representation $\overline{\Pi}(\bq)= [\Pi(\bq)]_{0,0}$ [Appendix~\ref{appen2} Eq.~\eqref{tavgA}].
We first proceed to obtain an approximate analytical result for the time-averaged free-particle polarization function. Two regimes of interest that exhibit qualitatively different behavior can be identified, {\it i.e.}, the weak drive regime with $\mathcal{V}_{2}<E_{F}<\homg$ and the strong drive regime where $\mathcal{V}_{2}>E_{F}$. In the following we discuss the weak drive regime, in which the Fermi level intersects with the zeroth order quasienergy band.
$\overline{\Pi}(\bq)$ is obtained by setting $n=0$ in Eq.~\eqref{diagonalpi}.
Since time-periodic driving leads to  the appearance of the Floquet sidebands, $\overline{\Pi}(\bq)$ contains two types of contributions: intraband terms with momentum transfers nesting within the same Floquet sideband's Fermi surface, and interband terms with momenta spanning the Fermi surfaces of distinct Floquet sidebands. To gain better insight into the distinct contributions comprising $\overline{\Pi}(\bq)$, we categorize $\overline{\Pi}(\bq)$ into intraband and interband contributions as follows,
\begin{equation}\label{piavg}
  \overline{\Pi}(\bq)=\overline{\Pi}^{{(0)}}_{{\rm intra}}(\bq)+\overline{\Pi}^{{(2)}}_{{\rm intra}}(\bq)+\overline{\Pi}^{{(2)}}_{{\rm inter}}(\bq)\;,
\end{equation}
where
\begin{equation}\label{intra0}
\overline{\Pi}^{{(0)}}_{{\rm intra}}(\bq)= \frac{1}{2\pi^2}\int d\bk \sum_{\alpha=\pm}\frac{\fne{0}}{\Ennq{0}{+\alpha\bq}{0}}, \\
\end{equation}
\begin{eqnarray}\label{intra2}
\overline{\Pi}^{{(2)}}_{{\rm intra}}(\bq)=\frac{1}{2\pi^2}&&\left(\frac{\mathcal{V}_{1}}{\homg}\right)^{2}\bigg[\int d\bk \sum_{\alpha,n=\pm} \frac{f(\epsilon_{n,\bk})k^2}{\Ennq{n}{+\alpha\bq}{n}}\nonumber\\
&&-2\int d\bk \sum_{\alpha=\pm}\frac{f(\epsilon_{0,\bk})(k^2+q^2)}{\Ennq{0}{+\alpha\bq}{0}}\bigg],
\end{eqnarray}
\begin{equation}\label{inter2}
\overline{\Pi}^{{(2)}}_{{\rm inter}}(\bq)=\frac{1}{2\pi^2}\left(\frac{\mathcal{V}_{1}}{\homg}\right)^{2}\int d\bk  \sum_{\alpha,n=\pm} \frac{f(\epsilon_{0,\bk})(q^2-k^2)}{\Ennq{n}{+\alpha\bq}{0}}\;.
\end{equation}
Note that there is no zeroth order interband contribution since they are unique to non-equilibrium. It becomes evident from Eqs.~\eqref{intra0}-\eqref{inter2} that the approximation up to second order in driving strength implies that the relevant Floquet bands with non-vanishing occupation are $n=0,\pm1$. The latter can also be explicitly seen from the time-averaged occupation
\begin{equation}\label{occup}
\overline{n}_{\bk}=-i\int\frac{d\omega}{2\pi}\left[\Giless_{\bk}(\omega)\right]_{0,0}\;,
\end{equation}
where $\left[\Giless_{\bk}(\omega)\right]_{0,0}$ is given in Appendix~\ref{appen4} with $n=0$ in Eq.~\eqref{glessapproximation}. Evaluating the above we find
\begin{equation}\label{occupfull}
\overline{n}_{\bk}=\fne{0}+\left(\frac{\mathcal{A}v_{k}}{2\Omega}\right)^{2}\bigg[\fne{-1}+\fne{1}-2\fne{0}\bigg],
\end{equation}
where $v_{k}=\hbar k/\ms$, and $\mathcal{A}$ is in Eq.~\eqref{Hmn}.

Returning to the free-particle polarization function, and integrating Eqs.~\eqref{intra0}-\eqref{inter2}, we have obtained the intraband and interband contributions in Eq.~\eqref{piavg} as
\begin{equation}\label{zero}
\overline{\Pi}^{{(0)}}_{{\rm intra}}(\bq)=\mathcal{D}\left[1-\Theta(q-2k_{0,F})\sqrt{1-\left(\frac{2k_{0,F}}{q}\right)^{2}}\right]\;,
\end{equation}
\begin{eqnarray}\label{one}
&&\overline{\Pi}^{{(2)}}_{{\rm intra}}(\bq)=\mathcal{D}\left(\frac{\mathcal{A}v_{q}}{2\Omega}\right)^{2}\bigg\{\\
&&-\frac{13}{6}+\left[\frac{2k^{2}_{0,F}+7q^{2}}{3q^{2}}\right]\sqrt{1-\left(\frac{2k_{0,F}}{q}\right)^{2}}\Theta(q-2k_{0,F})\nonumber\\
&&- \left[\frac{2k^{2}_{-1,F}+q^{2}}{6q^{2}}\right]\sqrt{1-\left(\frac{2k_{1,F}}{q}\right)^{2}}\Theta(q-2k_{1,F}) \bigg\}\;,\nonumber
\end{eqnarray}
\begin{eqnarray}\label{two}
&&\overline{\Pi}^{{(2)}}_{{\rm inter}}(\bq)=\mathcal{D}\left(\frac{\mathcal{A}v_{q}}{2\Omega}\right)^{2}\left\{\left[\frac{5}{3}-\left(\frac{\homg}{\epsilon_{0,q}}\right)^{2} \right]\right.\\
&&+\sum_{n=\pm 1}\frac{(q^{2}+2\ms n\Omega/\hbar)}{q^2}\sqrt{1-\zeta^2_{q}}\Theta\left(1-\zeta^2_{q}\right)\nonumber\\
&&\times\left.\left[-1 +\frac{(q^{2}+2\ms n\Omega/\hbar)^3}{q^4}\left(1+\frac{\zeta^2_{q}}{2}\right)\right]\right\}\;,\nonumber
\end{eqnarray}
here we have defined $v_{q}=\hbar q /\ms$, $\zeta_{q}=(2k_{0,F}q)/(q^{2}+2\ms n\Omega/\hbar)$, $\Theta$ is the Heaviside step function, $\mathcal{D}=\ms/(\pi\hbar^2)$ is the 2DES equilibrium density of states, $k_{n,F}=\sqrt{(E_{F}+n\homg-\mathcal{V}_{2})2\ms}/\hbar$ is the Fermi wave-vector associated with the Floquet mode $\epsilon_{n,k}$ [Eq.~\eqref{exactenergy}], and in this case we have considered $\mathcal{V}_{2}<E_{F}<\homg$. The well-known equilibrium free-particle polarization function is recovered by setting $\mathcal{A}=0$ in Eqs.~\eqref{zero}-\eqref{two}, since $\overline{\Pi}^{{(2)}}_{{\rm intra}}(\bq)=\overline{\Pi}^{{(2)}}_{{\rm inter}}(\bq)=0$, and the equilibrium form of the free-particle polarization function is given by Eq.~\eqref{zero}.
Here we note that, similar to the equilibrium case, the long-wavelength limit of the polarization function $\overline{\Pi}(\bq\rightarrow0)=\mathcal{D}$ since $\lim_{q \to 0} \overline{\Pi}^{{(2)}}_{{\rm inter}}(\bq)= \overline{\Pi}^{{(2)}}_{{\rm intra}}(\bq)=0$, and $\lim_{q \to 0}\overline{\Pi}^{{(0)}}_{{\rm intra}}(\bq)=\mathcal{D}$.

We have also performed exact numerical calculations of the time-average polarization function using Eqs.~\eqref{pi0finalnew}-\eqref{Gra} in the weak drive regime $\mathcal{V}_{2}<E_{F}<\homg$.  Fig.~\ref{fig1}(a)-(b) displays our numerically and analytically evaluated $\overline{\Pi}(\bq)$ for different driving strengths characterized by the dimensionless parameter $\mathcal{A}a$, where $\mathcal{A}=eE_{0}/(\hbar \Omega)$. An excellent agreement is seen between the approximate analytical results and the exact numerical results.
As is well known, the equilibrium polarization function for 2DESs exhibits a non-analyticity at $q = 2k_F$ ($k_F$ is the equilibrium Fermi wavevector) known as the Kohn anomaly. As we increase the driving strength,  $\overline{\Pi}(\bq)$ starts to deviate from its equilibrium value at $\mathcal{A}a = 0.1$, and the changes are clearly displayed when the driving strength reaches $\mathcal{A}a = 0.25$. We observe that the polarization function at a particular $q$ is decreased by increasing $\mathcal{A}a$; hence, screening effect on average is suppressed by periodic driving. Secondly, we notice that the Kohn anomaly feature corresponding to $q = 2k_F$ at equilibrium is shifted towards progressively smaller $q$ values as the driving strength is increased, accompanied by the emergence of kink features at other $q$ values. These features correspond to higher order Kohn anomalies arising from sharpness of the Fermi surface at intersecting Floquet sidebands.
The non-equilibrium behavior of the time-averaged polarization function exhibited in Fig.~\ref{fig1} can be understood from the approximate analytic results Eqs.~\eqref{zero}-\eqref{two}. The zeroth and second order intraband contributions Eqs.~\eqref{zero}-\eqref{one} reveal that the $n=0$ Kohn anomaly happens at $q=2k_{0,F}$, which decreases with increasing driving strength due to the irradiation generated energy shift $\mathcal{V}_2$.
In addition to the $n=0$ intraband Kohn anomaly, two new Kohn anomalies emerge at $q=2k_{1,F}$ and $q=k_{0,F}+k_{1,F}$. The former is an intraband Kohn anomaly while the latter is an interband Kohn anomaly. Both of these Kohn anomalies have contributions to  $\overline{\Pi}(\bq)$ that are proportional to $\mathcal{A}^2$,
hence becoming prominent at large values of $\mathcal{A}$ as illustrated in Fig.~\ref{fig1}. This observation is also associated with the increase in the higher order Floquet sidebands' occupation [Eq.~\eqref{occupfull}] with increasing $\mathcal{A}$. In addition to emergent Kohn anomalies induced by Floquet driving, $\overline{\Pi}(\bq)$ is suppressed by a ``background" non-equilibrium correction that is analytic, as shown in Eqs.~\eqref{zero}-\eqref{two}.

A qualitative change in $\overline{\Pi}(\bq)$ happens when $E_{F}<\mathcal{V}_{2}$ (strong drive regime). In this regime, the zeroth order Floquet sideband becomes unoccupied and $\overline{\Pi}^{{(0)}}_{{\rm intra}}(\bq)=\overline{\Pi}^{{(2)}}_{{\rm inter}}(\bq)=0$ [Eq.~\eqref{intra0} and Eq.~\eqref{inter2}]. Thus the contribution to the time-averaged free-particle polarization function arises uniquely from $\overline{\Pi}^{{(2)}}_{{\rm intra}}(\bq)$ yielding
\begin{eqnarray}\label{eflessef}
&&\overline{\Pi}(\bq)=-\mathcal{D}\left(\frac{\mathcal{A}v_{q}}{2\Omega}\right)^{2}\\
&&\times \left[\frac{2k^{2}_{-1,F}+q^{2}}{6q^{2}}\right]\sqrt{1-\left(\frac{2k_{1,F}}{q}\right)^{2}}\Theta(q-2k_{1,F})\;.\nonumber
\end{eqnarray}
The above equation indicates the absence of the $n=0$ intraband and interband contributions. Additionally, in this case, the time-averaged polarization function arises from the contributions of the $n=1$ [$\epsilon_{1,k}=\hbar^2\bk^2/(2\ms)+\mathcal{V}_{2}-\hbar\Omega$] sideband, since this is the only occupied band in the strong drive regime, as it can be deduced from Eq.~\eqref{occupfull} by setting $E_{F}<\mathcal{V}_{2}$. Consequently, the only non-analyticity displayed in this regime is the intraband Kohn anomaly associated with the $n=1$ sideband, {\it i.e}, $q=2k_{1,F}$.

We have found the matrix elements of the free-particle polarization function, Eqs.~\eqref{diagonalpi}-\eqref{pi3}. With the insights gained from the analysis of the time-averaged polarization function, we now proceed to find the time-averaged inverse dielectric function and the real-space representation of the screened impurity potential.

\subsection{Time-averaged Dielectric Function and Screened Impurity Potential}~\label{sec3d}
We have shown in  Sec.~\ref{sec3} that the  time-averaged screened potential of an immobile static impurity is given by Eq.~\eqref{screened2}, which is reproduced here for convenience:
 \begin{eqnarray}\label{screened11}
   \overline{V}_{{\rm sc}}(\bq)= \overline{\epsilon^{-1}}(\bq) \mathbb{V}_{{\rm C}}(\bq)\;,
\end{eqnarray}
here $\overline{\epsilon^{-1}}(\bq)$ is the time-averaged inverse dielectric function and $\mathbb{V}_{{\rm C}}(\bq)= -2\pi Z  e^2/(\epsilon_{b}\bq)$ is the bare impurity potential.  $\overline{\epsilon^{-1}}(\bq)$ can be obtained from the $(0,0)$ element of the inverse dielectric function matrix $\epsilon(\bq)$, which depends on the free-particle polarization function as (see Appendix~\ref{appen3})
\begin{equation}\label{newmateq}
\epsilon^{-1}(\bq)=\left[\mathbb{I}+v_{\rm C}(\bq)\Pi(\bq)\right]^{-1}\;,
\end{equation}
Substituting Eqs.~\eqref{diagonalpi}-\eqref{pi3} for $\Pi(\bq)$ into the above, performing the matrix inversion, and keeping terms up to $\mathcal{A}^2$
for consistency with our analytical approximation in Sec.~\ref{sec3b}, we find the time-averaged inverse dielectric function in the weak drive regime ($\mathcal{V}_{2}<E_{F}<\homg$)
\begin{eqnarray}\label{inverseespsilon1}
\overline{\epsilon^{-1}}(\bq)=&&\frac{1}{\left[1+v_{\rm C}(\bq)\Pi^{(0)}_{0,0}(\bq)\right ]} -\frac{\Pi^{(2)}_{0,0}(\bq)v_{\rm C}(\bq)}{\left[1+v_{\rm C}(\bq)\Pi^{(0)}_{0,0}(\bq)\right ]^2}\nonumber\\
&&+\frac{2 \Pi^{2}_{1,0}(\bq)v^2_{\rm C}(\bq)}{\left[1+v_{\rm C}(\bq)\Pi^{(0)}_{0,0}(\bq)\right ]^2\left[1+v_{\rm C}(\bq)\Pi^{(0)}_{1,1}(\bq)\right ]}\;.\nonumber\\
\end{eqnarray}
Here we have defined $\Pi^{(0)}_{0,0}(\bq)=\overline{\Pi}^{(0)}_{{\rm intra}}(\bq)$  where $\overline{\Pi}^{(0)}_{{\rm intra}}(\bq)$ is given in Eq.~\eqref{zero}, $\Pi^{(2)}_{0,0}(\bq)=\overline{\Pi}^{(2)}_{{\rm intra}}(\bq)+\overline{\Pi}^{(2)}_{{\rm inter}}(\bq)$, where $\overline{\Pi}^{(2)}_{{\rm intra/inter}}(\bq)$ are given in Eqs.~\eqref{one} and \eqref{two}, $\Pi_{1,0}(\bq)=ie^{i\theta_{q}}\left[\Pi(\bq)\right]_{1,0}$, $\left[\Pi(\bq)\right]_{1,0}$ is the $(1,0)$ matrix element of the free-particle polarization function defined in Eq.~\eqref{pi2}, and $\Pi_{1,0}(\bq)$ is explicitly given by
\begin{eqnarray}\label{pi10}
&&\Pi_{1,0}(\bq)=\\
&&\mathcal{D}\left(\frac{\mathcal{A}v_{q}}{2\Omega}\right)\left\{\left[1-\Theta(q-2k_{0,F})\sqrt{1-\left(\frac{2k_{0,F}}{q}\right)^{2}}\right]\right.\nonumber\\
&&\left.-\frac{1}{2}\sum_{n=\pm 1}\frac{(q^{2}+2\ms n\Omega/\hbar)}{q^2}\sqrt{1-\zeta^2_{q}}\Theta\left[1-\zeta^2_{q}\right]\right\}\;,\nonumber
\end{eqnarray}
where $\mathcal{D}$, $v_{q}$ and $\zeta_{q}$ are given in Eqs.~\eqref{zero}-\eqref{two}. The function $\Pi^{{(0)}}_{1,1}(\bq)$ is the driving strength independent part of the $\left[\Pi(\bq)\right]_{1,1}$ matrix element of the free-particle polarization function, given in Eq.~\eqref{diagonalpi}, and  $\Pi^{{(0)}}_{1,1}(\bq)$ is given by
\begin{equation}\label{pi110}
\Pi^{{(0)}}_{1,1}(\bq)= \frac{\mathcal{D}}{2}\sum_{n=\pm 1}\frac{(q^{2}+2\ms n\Omega/\hbar)}{q^2}\sqrt{1-\zeta^2_{q}}\Theta\left(1-\zeta^2_{q}\right)\;.
\end{equation}
In the absence of periodic driving $\mathcal{A}=0$, $\Pi^{{(2)}}_{0,0}(\bq)=\Pi_{1,0}(\bq) =0$ and the first term in \eqref{inverseespsilon1} recovers the equilibrium  inverse dielectric function~\cite{Rev2DEG} within RPA.

Similar to the time-averaged non-interacting polarization function, there is a qualitative change in the behavior of the time-averaged inverse dielectric function when   $E_{F}<\mathcal{V}_{2}$. If the energy shift caused by light irradiation exceeds the Fermi level, then $\Pi^{{(0)}}_{0,0}(\bq)=\Pi_{1,0}(\bq)=\Pi^{{(0)}}_{1,1}(\bq) =0$, and the only non-vanishing function in Eq.~\eqref{inverseespsilon1} is $\Pi^{{(2)}}_{0,0}(\bq)$, which is obtained in Eq.~\eqref{eflessef}. Hence, the time-averaged inverse dielectric function for the strong drive regime ($E_{F}<\mathcal{V}_{2}$) becomes
\begin{eqnarray}\label{inverseespsilon2}
\overline{\epsilon^{-1}}(\bq)=1 -\Pi^{{(2)}}_{0,0}(\bq)v_{\rm C}(\bq)\;.
\end{eqnarray}

We can gain some insight into the behavior of the screened potential Eq.~\eqref{screened11} in the long wavelength limit $\bq\rightarrow0$. In the weak drive regime $\mathcal{V}_{2}<E_{F}<\homg$, the inverse dielectric function \eqref{inverseespsilon1} becomes
\begin{equation}\label{epsTF1}
 \overline{\epsilon^{-1}}(\bq)\approx\frac{q}{q_{\rm TF}+q}\;,
\end{equation}
so that the time-averaged screened potential takes the form
\begin{eqnarray}\label{thomas1}
\overline{V}_{{\rm sc}}(\bq)\approx -\frac{Ze^2 (2\pi)}{\epsilon_{b}}\frac{1}{q +q_{\rm TF}}\;,
\end{eqnarray}
here $q_{\rm TF}=2\ms e^2/(\hbar^2 \epsilon_{b})$ is the Thomas-Fermi screening wavevector. We note that the latter two equations take the usual Thomas-Fermi forms of the screened Coulomb potential and the inverse dielectric function~\cite{Rev2DEG}. This is because the non-equilibrium contributions $\Pi^{(2)}_{0,0}(\bq)$, $\Pi^{{(0)}}_{1,1}(\bq)$, $\Pi_{1,0}(\bq)$ to the polarization function Eq.~\eqref{inverseespsilon1} vanish in the limit $\bq\rightarrow0$ due to their $q^2$ proportionality. In contrast, the equilibrium-like  contribution $\Pi^{{(0)}}_{0,0}(\bq)$ is proportional to the density of states $\mathcal{D}$ when $\bq \rightarrow0$. In the real space, the time-averaged screen potential takes the well-known large-distance asymptotic expression
$
 \overline{V}_{{\rm sc}}({\bm r})\approx -(Ze^2 q_{\rm TF})/[\epsilon_{b}(q_{\rm TF}r)^3]
$
in the Thomas-Fermi limit.

In the strong drive regime ($E_{F}<\mathcal{V}_{2}$) when the irradiation-induced quasienergy energy shift exceeds the Fermi level, we find that periodic driving has a dramatic effect on screening in the long wavelength limit. In this case the time-averaged inverse dielectric function is given by Eq.~\eqref{inverseespsilon2}, and with $\Pi^{{(2)}}_{0,0}(\bq\rightarrow0) = 0$ it becomes  $\overline{\epsilon^{-1}}(\bq)\approx 1$, \textit{i.e.}, the screening effect when averaged over time vanishes. Accordingly, the total potential in the system is $\overline{V}_{{\rm sc}}(\bq)\approx \mathbb{V}_{{\rm C}}(\bq)$ and its real space representation is
$
 \overline{V}_{{\rm sc}}({\bm r})\approx -(Ze^2)/(\epsilon_{b}r)
$.

The long-wavelength behavior of the inverse dielectric function leading to Thomas-Fermi screening is valid for momenta $q \ll q_{{\rm A,\min}}$, where $q_{{\rm A,\min}}$ denotes the smallest Kohn anomaly. The Thomas-Fermi approximation fails to describe the electron gas response to density perturbations that lead to large momentum transfer. For impurity screening the Thomas-Fermi screening is not a good approximation as it involves large momentum transfer processes. In particular, the non-analytic behavior in $\overline{\epsilon^{-1}}(\bq)$ cannot be neglected since the non-analyticities of $\overline{\epsilon^{-1}}(\bq)$ arising from the Kohn anomalies of the polarization function contribute predominantly to the screened potential $ \overline{V}_{{\rm sc}}({\bm r})$. To obtain the large-distance asymptotic behaviour of
\begin{equation}\label{realspecerep}
\overline{V}_{{\rm sc}}({\bm r})=\int\frac{d\bq e^{i\bq\cdot {\bm r}}}{(2\pi)^2}\overline{\epsilon^{-1}}(\bq)\mathbb{V}_{{\rm C}}(\bq)
\end{equation}
we make use of the {\it Riemann-Lebesgue} lemma which states that if a function oscillates rapidly around zero then the integral of this function
is small and the principal contribution to the integral arises from the behavior of the integrand in the vicinity of the non-analytic points~\cite{lemma,stern}. We provide the detailed derivation of the large-distance asymptotic limit of  Eq.~\eqref{realspecerep} in Appendix~\ref{appen5}. For the strong drive regime ($\mathcal{V}_{2}<E_{F}<\homg$) we find that
\begin{equation}\label{FinalFrid}
\overline{V}_{{\rm sc}}({\bm r})\approx\frac{Ze^2 q_{\rm TF}}{\epsilon_{b}}\sum_{q_{A} \in Q} \mathcal{R}^2[q_{A} , S(q_{A} )]\mathcal{F}_{q_{A}}(q_{\rm TF})\frac{\sin(q_{A}r)}{(q_{A}r)^2}\;,
\end{equation}
where $Q=\left\{2k_{0,F},2k_{1,F},k_{0,F}+k_{1,F}\right\}$ is the set of Kohn anomalies allowed in this energy regime,
$\mathcal{R}$,  $S$, and $\mathcal{F}_{q_{A}}$ are given in Appendix~\ref{appen5} [Eqs.~\eqref{R}, \eqref{S} and \eqref{fsum}]. From the asymptotic behaviour of the screened potential in Eq.~\eqref{FinalFrid}, we notice the appearance of Friedel oscillations. Interestingly, the irradiated 2DES displays multi-period Friedel oscillations. The new periods of Friedel oscillations arise from the Kohn anomalies associated with the occupied Floquet sidebands. On the other hand, the envelope of the oscillations decays as $r^{-2}$ as in equilibrium. This is due to the nature of the non-analyticities displayed in the inverse dielectric function, Eq.~\eqref{inverseespsilon1}, which contains a square root dependence similar to the equilibrium case [see Eqs.~\eqref{nona1}, \eqref{nona2}, and \eqref{nona3} in Appendix~\ref{appen5}].

\begin{figure}[t!]
  \centering
  \includegraphics[width=\columnwidth]{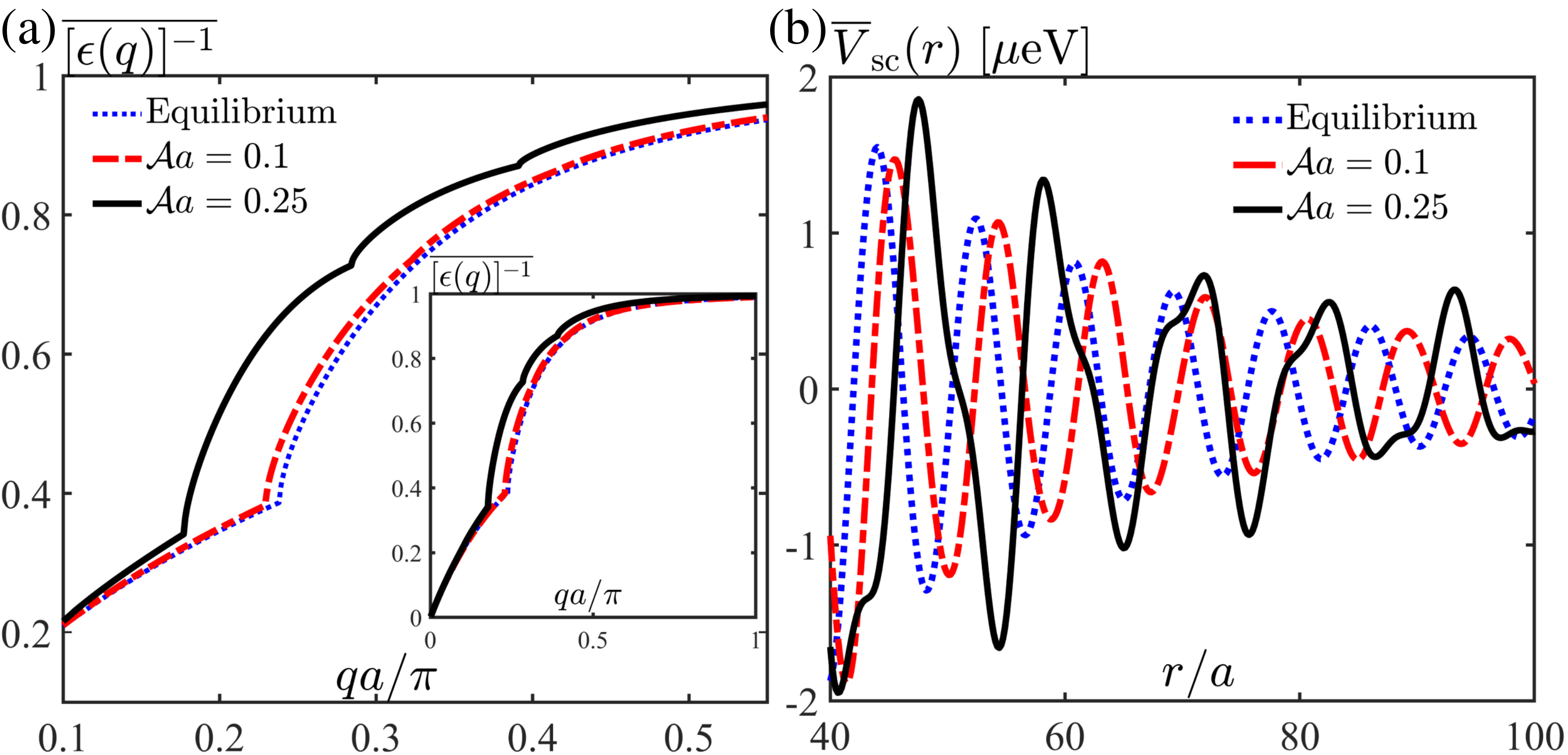}
  \caption{(a) Momentum-representation of the time-averaged inverse dielectric function and (b) asymptotic behaviour of the real space dependence of the screened potential displaying  Friedel oscillations, Eq.~\eqref{FinalFrid}. The light properties and the 2DES parameters are given in Fig.~\ref{fig1}. In this case $E_{F}>\mathcal{V}_{2}$ for all values of $\mathcal{A}$, and we have set $Z=1$ and the parameter $r_{s}=1/(\sqrt{\pi n_{c}}a^{*})=2$, where $a^{*}$ is the effective Bohr radius and $n_{c}$ is the carrier concentration~\cite{Rev2DEG}. The inset in (a) shows the an extended $q$-range of $\overline{\epsilon^{-1}}(\bq)$.}  \label{fig2}
\end{figure}

In the strong drive regime, $E_{F}<\mathcal{V}_{2}$, the inverse dielectric function undergoes a qualitative change, Eq. \eqref{inverseespsilon2}, and the real space representation of the screened potential becomes
\begin{equation}\label{2frid}
\overline{V}_{{\rm sc}}({\bm r})=\int_{0}^{\infty}\frac{q dqJ_{0}(qr)}{2\pi} \mathbb{V}_{{\rm C}}(q)\left[1 -\Pi^{{(2)}}_{0,0}(\bq)v_{\rm C}(q)\right].
\end{equation}
It is obvious that the first part of the integral above is simply the unscreened Coulomb potential, hence
\begin{equation}\label{2frid2}
\overline{V}_{{\rm sc}}({\bm r})=-\frac{Ze^2}{\epsilon_{b}r}-\int_{0}^{\infty}\frac{q dqJ_{0}(qr)}{2\pi} \mathbb{V}_{{\rm C}}(q)\Pi^{{(2)}}_{0,0}(\bq)v_{\rm C}(q).
\end{equation}
and the integral in the above equation can be evaluated by using the Riemann-Lebesgue lemma and following the steps in Appendix.~\ref{appen5}.
We then obtain the asymptotic limit of $\overline{V}_{{\rm sc}}({\bm r})$ as
\begin{eqnarray}\label{2frid3}
&&\overline{V}_{{\rm sc}}({\bm r})\approx\\
&&-\frac{Ze^2}{\epsilon_{b}r}+\frac{Ze^2 q_{\rm TF}}{\epsilon_{b}}\mathcal{R}^2[\bq,0]\mathcal{F}_{2k_{-1.F}}(0)\frac{\sin(2k_{1,F}r)}{(2k_{1,F}r)^2},\nonumber
\end{eqnarray}
where $\mathcal{R}$ and $\mathcal{F}$ are the same functions as in Eq.~\eqref{FinalFrid}. From Eq.~\eqref{2frid3} we notice that, for the irradiated 2DES, when the light matter-coupling strength is large enough to set $\mathcal{V}_{2}$ above $E_{F}$, the Coulomb potential becomes, on average, unscreened and accompanied by oscillations arising from the density-response of the electrons occupying higher Floquet sidebands.

Having found the analytical forms of the inverse dielectric function and the screened Coulomb potential in the periodically driven 2DES, we will proceed to discuss the main properties and features of these observables and their dependence on the driving field parameters.

\begin{figure}[t!]
  \centering
  \includegraphics[width=\columnwidth]{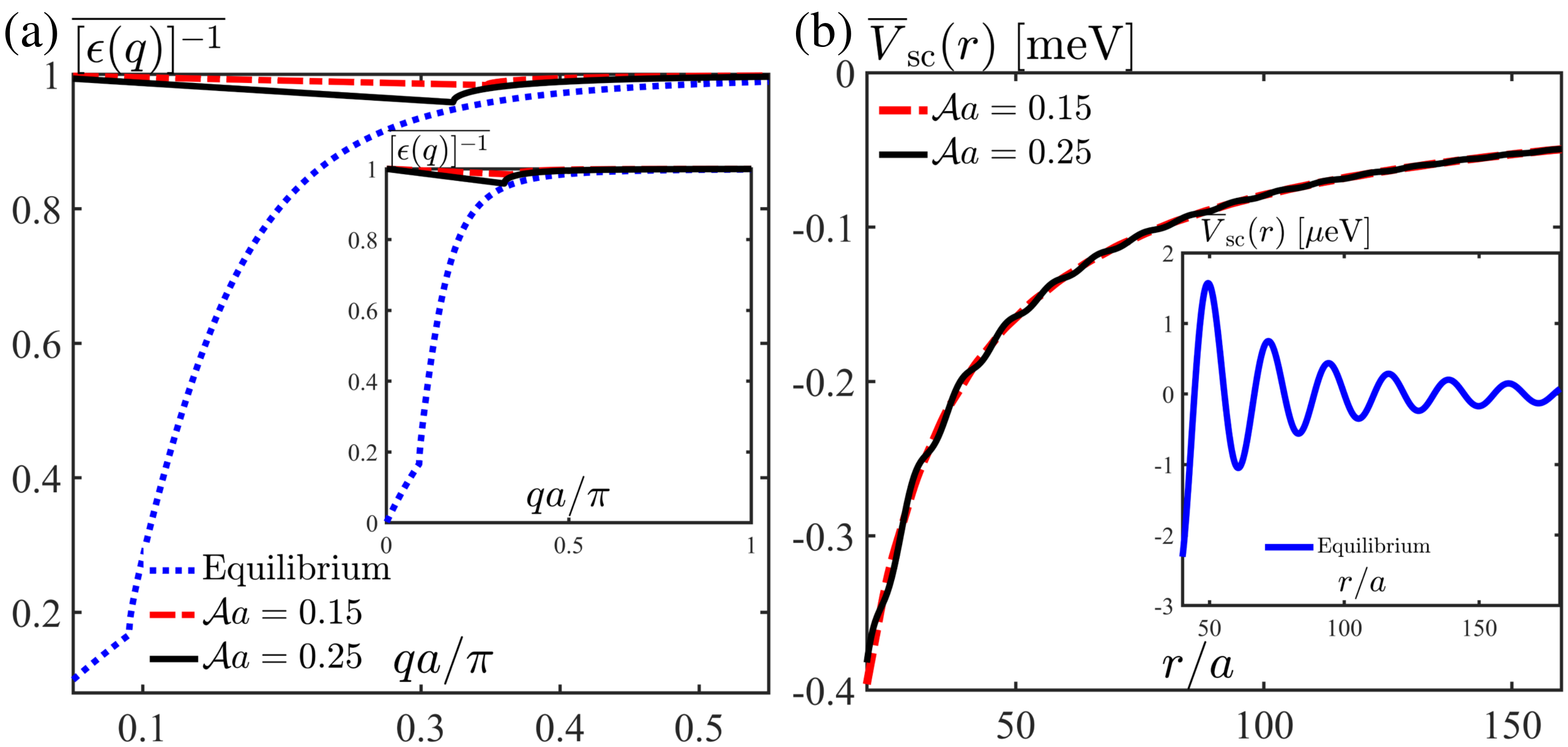}
  \caption{ (a) Time-averaged inverse dielectric function as a function of momentum for an irradiated system with $\mathcal{A}a=0.15, 0.25$, and equilibrium. The light parameters in this case are $E_{0}=120,210$ MV/m, $\homg$=0.3 eV and the Fermi level is set at $E_{F}=20$ meVs. All other 2DES parameters are similar to Fig.~\ref{fig1}. In this regime $E_{F}<\mathcal{V}_{2}$ for all values of $\mathcal{A}$. The inset in (a) shows $\overline{\epsilon^{-1}}(\bq)$ for a longer range $q$. As in Fig.~\ref{fig2} we have set $Z=1$ and $r_{s}=2$. (b) Real space representation of the ``screened" potential [Eq.~\eqref{2frid3}] showing the lack of the typical Friedel oscillations, and the inset shows the equilibrium real-space dependence of $V_{\rm sc}(r)$. }\label{fig3}
\end{figure}

\subsection{Periodic Driving Effects on the Screening Properties of 2DESs}\label{sec3e}

This section is devoted to discussing the effects of periodic driving on the inverse dielectric function and the screened Coulomb potential in a 2DES. We analyze two qualitatively different regimes distinguished by the relative location of the Fermi level $E_{F}$ and the light-induced band shift $\mathcal{V}_{2}$.

We first start with the weak drive regime where $\mathcal{V}_{2} <E_{F}<\hbar\Omega$. The time-averaged inverse dielectric function is given by Eq.~\eqref{inverseespsilon1} and is shown in Fig.~\ref{fig2}(a) for different driving strengths. At $q = 0$, the inverse dielectric function under periodic driving coincide with the equilibrium Thomas-Fermi limit as discussed in Eq.~\eqref{epsTF1}. The effects of periodic driving become pronounced at finite momenta. Fig.~\ref{fig2}(a) shows that increasing driving strength leads to the appearance of kink features corresponding to the out-of-equilibrium Kohn anomalies in the time-averaged polarization function (Fig.~\ref{fig1}). The intraband Kohn anomaly corresponding to the $n=0$ Floquet band is pronounced for all driving strength values.
We also note that periodic driving enhances $\overline{\epsilon^{-1}}(\bq)$ compared to static screening in equilibrium. This reduction in screening effectiveness arises because the time-averaged occupation number  Eq.~\eqref{occupfull} for the $n = 0$ Floquet band decreases with increasing band shift $\mathcal{V}_{2}$. Hence on average there are less electrons contributing to screening.

The presence of Kohn anomalies in the inverse dielectric function leads to the appearance of an oscillatory behavior in the screened potential, {\it i.e.}, Friedel oscillations~\cite{Friedel}. At equilibrium, the screened potential oscillates with a period $\pi/k_{F}$ given by the equilibrium Kohn anomaly $q=2k_{F}$  of the 2DES.
The spatial decay of these oscillations away from the impurity location goes as $r^{-2}$ in two-dimensions~\cite{Rev2DEG}. This case can be recovered by setting $\mathcal{A} = 0$ in our result  Eq.~\eqref{FinalFrid}.
When the system is illuminated with a monochromatic CP light, Eq.~\eqref{inverseespsilon1} and Fig.~\ref{fig2}(b), the occupation of the Floquet sidebands [see Eq.~\eqref{occupfull}] is reflected in the Friedel oscillations via the appearance of new irradiation generated periods indicating the emergence of new Kohn anomalies [Fig.~\ref{fig2}(a)]. These new periods modulate the dominant period arising from the $n=0$ Floquet mode. As the driving strength increases, the Floquet sidebands' occupation increases while the occupation of the $n=0$ band decreases, Eq.~\eqref{occupfull}. This enhances the contributions from  the intraband and interband Kohn anomalies at $q=2k_{1,F}$ and $k_{0,F}+k_{1,F}$ in Eq.~\eqref{FinalFrid} while decreases the  contribution at $2k_{0,F}$. All the Kohn anomalies share the same type of square root non-analyticity (see Sec.~\ref{sec3d}) and thus contribute to the same power law decay as $1/r^2$.

In the strong drive regime $\mathcal{V}_{2} >E_{F}$, a dramatic change in the inverse dielectric function is noticed in Fig.~\eqref{fig3}(a) when compared to equilibrium.
In addition to $\overline{\epsilon^{-1}}(\bq\rightarrow0) = 1$ in long wavelength limit, $\overline{\epsilon^{-1}}(\bq)$ remains close to $1$ for all values of $q$ indicating a suppression of averaged screening effect by periodic driving. The non-analytic feature is due to the $n=1$ intraband Kohn anomaly $q=2k_{1,F}$, corresponding to the time-averaged polarization function for this regime [Eq.~\eqref{eflessef}]. The absence of the additional Kohn anomalies in $\overline{\epsilon^{-1}}(\bq)$ is due to a depopulated $n=0$ Floquet mode that does not intersect $E_{F}$.
The qualitative change in the $\overline{\epsilon^{-1}}(\bq)$ leads to drastic changes in the screened potential Eq.~\eqref{2frid3} as shown in Fig.~\ref{fig3}(b). In contrast to the equilibrium case, the impurity potential becomes almost unscreened and modulated only by weak oscillations. When $\mathcal{V}_{2} >E_{F}$, the Fermi sea has been ``drained out'' of the principal quasienergy band $n = 0$.  Even though the Fermi level intersects the Floquet band $n=1$, the occupation of this band remains low since it is smaller by a factor of $\sim [\mathcal{A}v_{1,F}/(2\Omega)]^2 \ll 1$ (here $v_{1,F}$ is the Fermi velocity at the band) than the occupation of the $n =0$ Floquet band.

At equilibrium, embedding impurities in a 2DES leads to the impurity's potential screening and appearance of Friedel oscillations. Our work has shown that for an irradiated 2DES, if the driving strength yields an energy shift smaller than the Fermi energy, the Friedel oscillation in the 2DES will persist while having longer periods of oscillation modulated by secondary periods arising from the higher Floquet sidebands. However, we also noted that as the driving strength increases so that the optical energy shift becomes larger than the Fermi level, the external impurity potential becomes weakly screened due to the low occupation of the Floquet sidebands intersecting the Fermi level. Friedel oscillations in two-dimensional equilibrium systems, such as the Cu($111$) and Be(0001) surfaces, and graphene, have been observed employing scanning tunneling microscopy (STM)~\cite{stm1,stm2,stm3}. Thus the non-equilibrium behavior of Friedel oscillations we predicted could in principle be observed using STM on irradiated two-dimensional electron systems.

\section{Conclusion}\label{sec5}

We have studied the time-averaged screening response of a charged impurity in a two-dimensional electron system that is driven by monochromatic circularly polarized light. Screening properties including the non-equilibrium polarization function, inverse dielectric function and screened impurity potential are formally derived within the Floquet Green's function formalism. Our analysis of the time-averaged polarization function reveals additional Kohn anomalies, which arise from the Fermi level intersecting with the Floquet sidebands. The Floquet quasienergy picture allows us to identify two regimes, the weak and strong drive regimes, with drastically different screening responses that depend on Fermi level position and the light-induced Floquet band shift. We examined these two regimes with analytical calculations of the time-averaged polarization function up to second order of the driving strength, as well as the corresponding time-averaged inverse dielectric function and screened impurity potential. When the Floquet band shift is smaller than the Fermi level, we find that screening is accompanied by longer Friedel oscillation periods and the appearance of secondary periods that arise from the occupied Floquet sidebands intersecting the Fermi level. Non-equilibrium features, appearing in the polarization function as additional Kohn anomalies and in the screened impurity potential as additional Friedel oscillations, become more prominent as the driving strength is increased. This continues until the driving strength is large enough that the Floquet band shift becomes larger than the Fermi level. In this regime Friedel oscillations are replaced by a barely screened time-averaged  impurity potential modulated by very weak spatial modulations. The distinctive behavior in the screened impurity potential introduced by a periodic driving field could lead to a viable strategy for designing new Floquet-driven materials.


\acknowledgments{
This work was supported by the U.S. Department of Energy, Office of Science, Basic Energy Sciences under Early Career Award No.~DE-SC0019326.}


\appendix
\section{Fourier transform and inverse fourier transform of two-time periodic functions}\label{apend1}
The Fourier transform of a two-time periodic function in the Floquet representation is~\cite{mills,ono}
\begin{eqnarray}\label{att}
\ma&&(t,t')=\\
&&\sum_{m}\int_{-\infty}^{\infty} \frac{d\omega}{2\pi}e^{-i[\omega+(m+n)\Omega/2] \tau} e^{-i(m-n)\Omega T}[\ma(\omega)]_{m,n},\nonumber
  \end{eqnarray}
and the transform of a two-time function convolution is
  \begin{eqnarray}\label{att2}
  &&\ma(t,t')= \int d t''\mb(t,t'')\mc (t'',t')=\sum_{m,j} \int_{-\infty}^{\infty}\frac{d\omega}{2\pi}\bigg\{\\
   &&e^{-i[\omega+(m+n)\Omega/2] \tau} e^{-i(m-n)\Omega T} [\mb(\omega) ]_{m,j}[\mc(\omega)]_{j,n}\bigg\},\nonumber
\end{eqnarray}
The Floquet indices $m$, $j$ and $n$ take the values $\{0, \pm 1, \pm2, \ldots \}$, $-\infty<\omega<\infty$ is the extended zone frequency, and the inverse transforms of Eqs.~\eqref{att} and \eqref{att2} are
\begin{eqnarray}\label{att3}
&&\left[\ma (\omega) \right]_{m,n}=\\
&&\int_{0}^{\mathcal{T}}\frac{dT}{\mathcal{T}}\int_{-\infty}^{\infty}d\tau e^{i[\omega +(m+n)\Omega/2 ]\tau}e^{i(m-n)\Omega T}\ma(t,t'),\nonumber
\end{eqnarray}
and
\begin{eqnarray}\label{att4}
&&\left[\ma (\omega) \right]_{m,n}=\sum_{j} [\mb(\omega) ]_{m,j}[\mc(\omega)]_{j,n} = \int_{0}^{\mathcal{T}}\frac{dT}{\mathcal{T}}\int_{-\infty}^{\infty}d\tau\bigg\{\nonumber\\
&& e^{i[\omega +(m+n)\Omega/2 ]\tau} e^{i(m-n)\Omega T}\bigg\}\int_{-\infty}^{\infty} d t''\mb(t,t'')\mc (t'',t'),
\end{eqnarray}
respectively.
\section{Time-averaged Observables}\label{appen2}
 In order to find the time-average of a two-time periodic function $\ma(t,t')$ we recall that its Wigner transform is~\cite{flreviewOKA,flreviewOKA2}
\begin{equation}\label{wigtavg}
\ma(\omega, T)=\int_{-\infty}^{\infty}d\tau e^{i\omega \tau} \ma(t,t')\;,
\end{equation}
then we substitute the Fourier transform in Eq.~\eqref{att}, integrate over $\tau$ and $\omega$ to arrive at
\begin{equation}\label{wigtavg2}
\ma(\omega, T)= \sum_{m}e^{-i(m-n)\Omega T}\left[\ma\left(\omega-\frac{(m+n)}{2}\Omega\right)\right]_{m,n} \;.
\end{equation}
Averaging the function above over one cycle, {\it i.e.} time-averaging,
\begin{eqnarray}\label{wigtavg3}
\overline{\ma}(\omega)&&=\\
&& \int_{0}^{\mathcal{T}}\frac{dT}{\mathcal{T}}\sum_{m}e^{-i(m-n)\Omega T}\left[\ma\left(\omega-\frac{(m+n)}{2}\Omega\right)\right]_{m,n},\nonumber
\end{eqnarray}
reveals that the time average of the two-time periodic function, $\ma(t,t')$, is simply the $(0,0)$-Floquet component,
\begin{equation}\label{tavgA}
\overline{\ma}(\omega)=\left[\ma\left(\omega\right)\right]_{0,0} \;.
\end{equation}

\section{Screened Impurity Potential}\label{appen3}
 Considering an immobile impurity that remains static against light irradiation (see Sec.~\ref{sec3}). Within these impurity considerations and with the aid of Eq.~\eqref{att} the screened potential in Eqs.~\eqref{sc1}-\eqref{eps2} can be written as
\begin{eqnarray}
  &&\Phi_{{\rm sc}}(\bq,t)=\\
   &&\Phi_{{\rm ext}}(\bq)\sum_{n}\int_{-\infty}^{\infty}\frac{dt' d\omega}{(2\pi)}[\epsilon^{-1}(\bq,\omega)]_{n,0}e^{-i\omega (t-t')}e^{-in\Omega t},\nonumber
\end{eqnarray}
integrating over $t'$ and over $\omega$,
\begin{eqnarray}
  \Phi_{{\rm sc}}(\bq,t)= \Phi_{{\rm ext}}(\bq)\sum_{n}[\epsilon^{-1}(\bq,0)]_{n,0}e^{-in\Omega t}\;.
\end{eqnarray}
Moreover, averaging the screened potential over its period, {\it i.e.} $\int_{0}^{\mathcal{T}}dt/\mathcal{T}$, and recalling that $V(\bq)=-e\Phi(\bq)$, we get the time-averaged screened Coulomb interaction,
\begin{eqnarray}\label{screened1}
  \overline{V}_{{\rm sc}}(\bq)= [\epsilon^{-1}(\bq,0)]_{0,0}\mathbb{V}_{{\rm C}}(\bq)\;,
\end{eqnarray}
where $\mathbb{V}_{{\rm C}}=-Z v_{{\rm C}}(\bq)$, $v_{{\rm C}}(\bq)$ is given in Eq.~\eqref{eps1}. Substituting Eq.~(\ref{matforms}c) in  Eq.~(\ref{matforms}b), we can express the time averaged screened Coulomb potential, Eq.~\eqref{screened1}, as
\begin{eqnarray}\label{screened2a}
  \overline{V}_{{\rm sc}}(\bq)= \left[\mathbb{I}+ v_{\rm C}(\bq)\Pi(\bq)\right]^{-1}_{0,0}\mathbb{V}_{{\rm C}}(\bq)\;.
\end{eqnarray}

\section{Approximate Forms of Floquet Green’s Functions}\label{appen4}

 In this appendix we provide the approximate forms of $\Giret_{\bk}(\omega)$, $\Giadv_{\bk}(\omega)$, and $\Giless_{\bk}(\omega)$ within our approximate scheme. In the interest of obtaining the analytic dependence of $\Pi(\bq)$ we take $\eta\rightarrow0$. Considering $(\mathcal{V}_{1}k/\homg)^{n}\approx 0$ for $n>2$, a close inspection of $\Giret_{\bk}(\omega)$ and $\Giadv_{\bk}(\omega)$ for increasing matrix dimensions, {\it e.g.}, $5\times5 \dots 13\times 13$, shows that these matrices have a convergent pentadiagonal form allowing for their analytical closed form
\begin{eqnarray}\label{GRAexact}
  &&[{\rm G}^{{\rm R/A}}_{\bk}(\omega)]_{n,m} =\left[\gra_{n}-2(\mvk)^2 \prod^{n+1}_{l=n-1}\gra_{l}\right]\delta_{n,m} \\
  &&+\mvk\left\{\left[ie^{i\theta_{k}} \prod_{l=n-1}^{n}\gra_{l}\delta_{n,m+1}-ie^{-i\theta_{k}} \prod_{l=n}^{n+1}\gra_{l}\delta_{n+1,m}\right]\right.\nonumber\\
   &&+\mvk\left.\left[ e^{2i\theta_{k}} \prod_{l=n-2}^{n}\gra_{l}\delta_{n,m+2}+e^{-2i\theta_{k}}\prod_{l=n}^{n+2}\gra_{l}\delta_{n+2,m}\right]\right\} \nonumber,
\end{eqnarray}
where we have defined, $\gra_{n}=(\omega \pm i\eta -\epsilon_{\bk ,n})^{-1}$, $\epsilon_{n,\bk}=Ak^2+\mathcal{V}_{2}-n\homg$ is the quasienergy corresponding to the $n^{\rm th}$ Floquet band, Eq.~\eqref{exactenergy}. With the analytical forms of $\Giret_{\bk}(\omega)$ and $\Giadv_{\bk}(\omega)$, and  up to order $(\mathcal{V}_{1}k/\homg)^2$, we deduce the form of $\Giless_{\bk}(\omega)=2i\eta\Giret_{\bk}(\omega)\mathcal{F}(\omega)\Giadv_{\bk}(\omega)$, where $\mathcal{F}(\omega)$ is a diagonal matrix, such that, $[\mathcal{F}(\omega)]_{n,m}=f(\omega+n\homg)\delta_{n,m}$ and $f(\omega)=\Theta(E_{F}-\omega)$ is the zero temperature Fermi function. Then up to second order in the light driving strength we find that, similar to ${\rm G}^{{\rm R/A}}_{\bk}(\omega)$, $\Giless_{\bk}(\omega)$ is pentadiagonal and given by
\begin{eqnarray}
 &&[\Giless_{\bk} (\omega)]_{n,n}=2\pi i\del{n}\left\{\fn{n}+(\mvk)^{2}\Bigg[\nonumber\right.\\
 &&\left.\left.\frac{2\fn{n}}{\Ene{n-1}\Ene{n+1}}+\sum_{\alpha=\pm1}\frac{\fnt{(n+\alpha)}}{\En{n+\alpha}{2}}\right]\right\},\nonumber\\
 &&{}[\Giless_{\bk} (\omega)]_{n+1,n}= 2\pi \mvk e^{-i\theta_{k}}\bigg\{\nonumber\\
 &&\;\;\;\;\;\;\;\;\;\sum_{\alpha=0,1}\frac{\fnt{(n+1-\alpha)}\del{n+1-\alpha}}{\Ene{n+\alpha}}\bigg\},\nonumber
 \end{eqnarray}
 \begin{eqnarray}
  &&{}[\Giless_{\bk} (\omega)]_{n+2,n}=-2\pi i (\mvk)^{2}e^{-2i\theta_{k}}\bigg[\nonumber\\
  &&+\sum_{\alpha=0}^{2}\frac{\fnt{(n+\alpha)}\del{n+\alpha}}{\Ene{n+2-2\lfloor\alpha/2\rfloor}\Ene{n+\alpha-(-1)^{\alpha+1}}}\bigg], \nonumber\\
  &&{}[\Giless_{\bk} (\omega)]_{n,n+1}=-[\Giless_{\bk} (\omega)]^{*}_{n+1,n}\;,\nonumber\\
  && [\Giless_{\bk} (\omega)]_{n,n+2}=-[\Giless_{\bk} (\omega)]^{*}_{n+2,n}\;,\label{glessapproximation}
\end{eqnarray}
where $\lfloor x \rfloor ={\rm max} \left\{n\in Z | n\le x\right\}$ is the floor function.

\section{Asymptotic Behaviour of the Screened Coulomb Potential}\label{appen5}

In this appendix we find the long-distance asymptotic limit of Eq.~\eqref{realspecerep} for $\mathcal{V}_{2}<E_{F}<\homg$. We start by integrating the angular dependence of  Eq.~\eqref{realspecerep} with the aid of the Jacoby-Anger expansion, and arrive at
\begin{eqnarray}\label{frid1}
&&\overline{V}_{{\rm sc}}({\bm r})=\\
&&\bigintsss_{0}^{\infty}\left\{\frac{\mathbb{V}_{{\rm C}}(\bq)}{\left[1+v_{\rm C}(\bq)\Pi^{{(0)}}_{0,0}(\bq)\right ]} -\frac{\Pi^{{(2)}}_{0,0}(\bq)\mathbb{V}_{{\rm C}}(\bq)v_{\rm C}(\bq)}{\left[1+v_{\rm C}(\bq)\Pi^{{(0)}}_{0,0}(\bq)\right ]^2}\right.\nonumber\\
&&\left.+\frac{2 \Pi^{2}_{1,0}(\bq)\mathbb{V}_{{\rm C}}(\bq)v^2_{\rm C}(\bq)}{\left[1+v_{\rm C}(\bq)\Pi^{{(0)}}_{0,0}(\bq)\right ]^2\left[1+v_{\rm C}(\bq)\Pi^{{(0)}}_{1,1}(\bq)\right ]}\right\}q dq J_{0}(q r)\;.\nonumber
\end{eqnarray}
To illustrate the procedure of finding the asymptotic form of $\overline{V}_{{\rm sc}}({\bm r})$ we will explicitly find the  asymptotic limit for the first integral comprising Eq.~\eqref{frid1}, and follow a similar procedure for the remaining two parts. Starting with
\begin{eqnarray}\label{frid2}
&&\overline{V}_{{\rm sc},1}({\bm r})=\bigintsss_{0}^{\infty}{\frac{q dq J_{0}(q r)\mathbb{V}_{{\rm C}}(\bq)}{\left[1+v_{\rm C}(\bq)\Pi^{{(0)}}_{0,0}(\bq)\right ]}}\;,
\end{eqnarray}
we identify that the non-analyticity present in the integrand is displayed at $q=2k_{0,F}$ and arises from Kohn anomaly in $\Pi^{{(0)}}_{0,0}(\bq)$. Then by considering that $2k_{0,F}r\gg 1$ we write the integral in the vicinity of the non-analytic point, such that
\begin{eqnarray}\label{frid3}
&&\overline{V}_{{\rm sc},1}({\bm r})\approx\\
&&\mathcal{L}(2k_{0,F})\int_{2k_{0,F}-\varsigma}^{2k_{0,F}+\varsigma}\frac{\delta\left[\Pi^{{(0)}}_{0,0}(\bq,2k_{0,F})\right]\cos(qr-\pi/2) }{\sqrt{\pi k_{0,F}r }}\;,\nonumber
\end{eqnarray}
where we have defined
\begin{equation}\label{L}
\mathcal{L}(x)=-\frac{x\mathbb{V}_{{\rm C}}(x)v_{\rm C}(x)}{(2\pi )\left[1+v_{\rm C}(x)\Pi^{{(0)}}_{0,0}(x)\right ]^2}\;,
\end{equation}
and
\begin{equation}\label{delta}
\delta\left[F(\bq,x)\right]=F(\bq\approx x )-F(x)\;,
\end{equation}
where $F$ is an arbitrary function of $\bq$, then
\begin{equation}\label{nona1}
\delta\left[\Pi^{{(0)}}_{0,0}(\bq,2k_{0,F})\right]=-\mathcal{D}\sqrt{\frac{q-2k_{0,F}}{k_{0,F}}}\Theta(q-2k_{0,F})\;,
\end{equation}
to obtain the latter equation we have used the definition in Eq.~\eqref{zero} and \eqref{inverseespsilon1}. Moreover $\varsigma$ in Eq.~\eqref{frid3} is a small positive number, $\varsigma\ll2k_{0,F}$. Then, integrating Eq.~\eqref{frid3} by parts, and changing variables to $\tilde{q}=\sqrt{q-2k_{0,F}}$, we get
\begin{eqnarray}\label{frid4}
&&\overline{V}_{{\rm sc},1}({\bm r})\approx\\
&&\frac{2\mathcal{L}(2k_{0,F})\mathcal{D}}{\sqrt{\pi}k_{F}}\frac{1}{r^{3/2}}\int_{0}^{\infty}d\tilde{q}\sin(\tilde{q}^2r+2k_{0,F}r-\pi/4)\;,\nonumber
\end{eqnarray}
we have extended the latter integral to $\infty$ as the main contribution of the integral arises from $\tilde{q}=0$. Then we perform the integral in Eq.~\eqref{frid4} to get
\begin{eqnarray}\label{frid5}
&&\overline{V}_{{\rm sc},1}({\bm r})\approx \frac{Ze^2 q_{\rm TF}}{\epsilon_{b}}\mathcal{R}^2[2k_{0,F}, S(2k_{0,F})]\frac{\sin(2k_{0,F}r)}{(2k_{0,F}r)^2}\;,\nonumber\\
\end{eqnarray}
where we have defined
\begin{equation}\label{R}
  \mathcal{R}[x, S(x)]=\frac{x}{x +q_{\rm TF}S(x)}\;,
\end{equation}
and
\begin{equation}\label{S}
S(x)=1-\sqrt{1-(2k_{0,F}/x)^2}\;.
\end{equation}
Proceeding to the second part of the integral in Eq.~\eqref{frid1}, {\it i.e.},
\begin{equation}\label{frid21}
\overline{V}_{{\rm sc},2}({\bm r})= -\bigintsss_{0}^{\infty}\frac{q dq J_{0}(q r)\Pi^{{(2)}}_{0,0}(\bq)\mathbb{V}_{{\rm C}}(\bq)v_{\rm C}(\bq)}{\left[1+v_{\rm C}(\bq)\Pi^{{(0)}}_{0,0}(\bq)\right ]^2}\;,
\end{equation}
 we recognize that the integrand contains three non-analytic points arising from the Kohn anomalies of $\Pi^{{(2)}}_{0,0}(\bq)$ and $\Pi^{{(0)}}_{0,0}(\bq)$ which are located at $q_{A}=2k_{0,F},2k_{1,F}$ and $k_{0,F}+k_{1,F}$. Then by considering $q_{A}r\gg 1$, the asymptotic expression of the integral in Eq.~\eqref{frid21} can be recast into
\begin{eqnarray}\label{frid22}
&&\overline{V}_{{\rm sc},2}({\bm r})\approx\\
&&\sum_{q_{A} \in Q}\left\{\mathcal{L}(q_{A})\int_{q_{A}-\varsigma}^{q_{A}+\varsigma}\frac{\delta\left[\Pi^{{(2)}}_{0,0}(\bq,q_{A})\right]\cos(qr-\pi/2)}{\sqrt{\pi q_{A}r/2 }}\right\}\nonumber\\
&&+\frac{2\mathcal{L}(2k_{0,F})v_{\rm C}(2k_{0,F})\Pi^{{(2)}}_{0,0}(2k_{0,F})}{[1+v_{\rm C}(2k_{0,F})\Pi^{{(0)}}_{0,0}(2k_{0,F})]}\Bigg\{\nonumber\\
&&\left.\int_{2k_{0,F}-\varsigma}^{2k_{0,F}+\varsigma}\frac{\delta\left[\Pi^{{(0)}}_{0,0}(\bq,2k_{0,F})\right]\cos(qr-\pi/2) }{\sqrt{\pi k_{0,F}r }}\right\}\;,\nonumber
\end{eqnarray}
where $Q=\left\{2k_{0,F},2k_{1,F},k_{0,F}+k_{1,F}\right\}$,
\begin{equation}\label{nona2}
\delta\left[\Pi^{{(2)}}_{0,0}(\bq,2k_{0,F})\right] = \frac{5}{2}\mathbb{G}(2k_{0,F})\;,
\end{equation}
\begin{equation}
\delta\left[\Pi^{{(2)}}_{0,0}(\bq,2k_{1,F})\right] = -\mathbb{G}(2k_{0,F})\;,\nonumber
\end{equation}
\begin{eqnarray}
\delta\left[\Pi^{{(2)}}_{0,0}(\bq,2k_{1,F})\right]&&=-\mathbb{G}(q_{0,1})\nonumber\\
&&\times\frac{\sqrt{2k_{0,F}k_{1,F}}}{(q_{0,1})}\left[2-3\left(\frac{2k_{0,F}}{q_{0,1}}\right)^2\right],\nonumber
\end{eqnarray}
we have used the definitions in Eqs.~\eqref{one},~\eqref{two},~\eqref{inverseespsilon1} and \eqref{delta} to find the equations above. We also defined $q_{0,1}=k_{0,F}+k_{1,F}$, and
\begin{equation}\label{G}
\mathbb{G}(x)=\mathcal{D}\left(\frac{\mathcal{A}v_{q_{A}}}{2\Omega}\right)^2\sqrt{\frac{q-x}{x}}\Theta(q-x)\;.
\end{equation}
By following similar steps to Eqs.~\eqref{frid3}-\eqref{frid5} we find the asymptotic behaviour of $\overline{V}_{{\rm sc},2}({\bm r})$ in Eq.~\eqref{frid21},
\begin{equation}\label{frid23}
\overline{V}_{{\rm sc},2}({\bm r})\approx \frac{Ze^2 q_{\rm TF}}{\epsilon_{b}}\sum_{q_{A} \in Q} \mathcal{R}^2[q_{A} , S(q_{A} )]\mathcal{F}_{1,q_{A}}(q_{\rm TF})\frac{\sin(q_{A}r)}{(q_{A}r)^2},
\end{equation}
where $\mathcal{R}$ and $S$ are given in Eqs.~\eqref{R} and Eq.~\eqref{S}, and  $\mathcal{F}_{1,q_{A}}(q_{\rm TF})$ are given by
\begin{eqnarray}\label{f1}
&&\mathcal{F}_{1,2k_{0,F}}(q_{\rm TF})=-\left(\frac{\mathcal{A}v_{k_{0,F}}}{\Omega}\right)^2\bigg\{\\
&&\left.\frac{5}{2}-\frac{q_{\rm TF}\mathcal{R}[2k_{0,F},S(2k_{0,F})]}{k_{0,F}}\left[\frac{1}{2}+\frac{5}{3}\left(\frac{\homg}{\epsilon_{0,2k_{0,F}}}\right)^2\right]\right\},\nonumber\\
&&\mathcal{F}_{1,q_{0,1}}(q_{\rm TF})=\left(\frac{\mathcal{A}v_{q_{0,1}}}{2\Omega}\right)^2\sqrt{\frac{k_{0,F}k_{0,-1}}{q^{2}_{0,-1}}}\left[2-3\left(\frac{2k_{0,F}}{q_{0,1}}\right)^2\right],\nonumber\\
&&\mathcal{F}_{1,2k_{1,F}}(q_{\rm TF})=\left(\frac{\mathcal{A}v_{k_{1,F}}}{\Omega}\right)^2\;.\nonumber
\end{eqnarray}
The third integral of Eq.~\eqref{frid1},
\begin{eqnarray}\label{frid31}
\overline{V}_{{\rm sc},3}&&({\bm r})=\\
&&-\bigintsss_{0}^{\infty}\frac{2q dq J_{0}(q r) \Pi^{2}_{1,0}(\bq)\mathbb{V}_{{\rm C}}(\bq)v^2_{\rm C}(\bq)}{\left[1+v_{\rm C}(\bq)\Pi^{{(0)}}_{0,0}(\bq)\right ]^2\left[1+v_{\rm C}(\bq)\Pi^{{(0)}}_{1,1}(\bq)\right ]}.\nonumber
\end{eqnarray}
The integrand in the latter equation has non-analytic behaviour at $q_{A}=2k_{0,F}$ and $k_{0,F}+k_{1,F}=q_{0,1}$ since $\Pi_{1,0}(\bq)$, $\Pi^{{(0)}}_{1,1}(\bq)$ and $\Pi^{{(0)}}_{0,0}(\bq)$ contain Kohn anomalies at $q_{A}$. Similar to the previous two cases we consider the asymptotic limit $q_{A}r\gg1$ and we recast Eq.~\eqref{frid31} into
\begin{eqnarray}
&&\overline{V}_{{\rm sc},3}({\bm r})\approx\\
&&\sum_{q_{A} \in Q'} \mathcal{L}(q_{A})\Upsilon(q_{A})\left\{ \int_{q_{A}-\varsigma}^{q_{A}+\varsigma}\frac{\delta\left[\Pi_{1,0}(\bq,q_{A})\right]\cos(qr-\pi/2)}{\sqrt{\pi q_{A}r/2}}\right.\nonumber\\
&&\left.+\tau(q_{A})\Upsilon(q_{A}) \int_{q_{A}-\varsigma}^{q_{A}+\varsigma}\frac{\delta\left[\Pi^{{(0)}}_{1,1}(\bq,q_{A})\right]\cos(qr-\pi/2)}{\sqrt{\pi q_{A}r/2}}\right\}\;,\nonumber
\end{eqnarray}
where $Q'=\left\{2k_{0,F}, q_{0,1} \right\}$, and we have defined,
\begin{eqnarray}
  &&\Upsilon(x) =-\frac{2v_{\rm C}(x)\Pi_{1,0}(x)}{[1+v_{\rm C}(x)\Pi^{{(0)}}_{1,1}(x)]}  \nonumber\\
 &&\tau(x)=\begin{cases} 1/2 &\mbox{if } x=q_{0,1} \\
               \frac{[1+v_{\rm C}(x)\Pi^{{(0)}}_{1,1}(x)]}{2[1+v_{\rm C}(x)\Pi^{{(0)}}_{0,0}(x)]}& \mbox{if } x= 2k_{0,F} \end{cases}
\end{eqnarray}
and we also note that $\Upsilon(2k_{0,F})=0$, since $\Pi_{1,0}(2k_{0,F})$ as it can be deduced from Eq.~\eqref{pi10} by direct substitution. Hence, the contribution of $q=2k_{0,F}$ to  Eq.~\eqref{frid31} is zero. Additionally $\Pi^{{(0)}}_{1,1}(q_{0,1})=\mathcal{D}$, and
\begin{eqnarray}\label{nona3}
&&\delta\left[\Pi_{1,0}(\bq,q_{0,1})\right]=\mathbb{G}(q_{0,1})/\left[\mathcal{A}q_{0,1}/(2\Omega)\right]\;,\\
&&\delta\left[\Pi^{{(0)}}_{1,1}(\bq,q_{0,1})\right]=-\delta\left[\Pi_{1,0}(\bq,q_{0,1})\right]/\left[\mathcal{A}q_{0,1}/(2\Omega)\right].\nonumber
\end{eqnarray}
Following the steps in Eqs.~\eqref{frid3}-\eqref{frid5}, we find
\begin{eqnarray}\label{frid32}
&&\overline{V}_{{\rm sc},3}({\bm r})\approx\\
&&\frac{Ze^2 q_{\rm TF}}{\epsilon_{b}}\mathcal{R}^2[q_{0,1} , S(q_{0,1} )]\mathcal{F}_{2,q_{0,1}}(q_{\rm TF})\frac{\sin(q_{0,1}r)}{(q_{0,1}r)^2}\;, \nonumber
\end{eqnarray}
where
\begin{eqnarray}\label{f2}
&&\mathcal{F}_{2,q_{0,1}}(q_{\rm TF})=\left(\frac{\mathcal{A}v_{q_{0,1}}}{2\Omega}\right)^2\Bigg\{\\
 &&2\left[\frac{q_{\rm TF}}{q_{0,1}}\right]\mathcal{R}[q_{0,1},S(2k_{0,F})]S(q_{0,1})\nonumber\\
&&\left. \times\left[\left(\frac{q_{\rm TF}}{q_{0,1}}\right)\mathcal{R}[q_{0,1},S(2k_{0,F})]S(q_{0,1})-2\right]\right\}\;.\nonumber
\end{eqnarray}

Then we have found the asymptotic form of $V_{\rm sc}(r)$ in Eq.~\eqref{FinalFrid}, where in this equation we have defined,
\begin{eqnarray}\label{fsum}
&&\mathcal{F}_{2k_{0,F}}(q_{\rm TF})=1+\mathcal{F}_{1,2k_{0,F}}(q_{\rm TF})\;, \\
&&\mathcal{F}_{q_{0,1}}(q_{\rm TF})=\mathcal{F}_{1,q_{0,1}}(q_{\rm TF})+\mathcal{F}_{2,q_{0,1}}(q_{\rm TF})\;, \nonumber\\
&&\mathcal{F}_{2k_{1,F}}(q_{\rm TF})=\mathcal{F}_{1,2k_{1,F}}(q_{\rm TF})\nonumber\;,
\end{eqnarray}
and the functions $\mathcal{F}_{1/2,{q_{A}}}(q_{\rm TF})$ are given in Eqs.~\eqref{f1} and \eqref{f2}.

\bibliographystyle{apsrev4-2}
\bibliography{refs2}
\end{document}